\documentclass[default]{aastex7}

\newcommand{\lynx}{\texttt{LightCurveLynx}}
\newcommand{\ztfsn}{ZTFSNDR2}
\newcommand{\pzflow}{\texttt{PZFlow}}
\newcommand{\sncosmo}{\texttt{SNCosmo}}
\newcommand{\redback}{\texttt{Redback}}

\usepackage{amsmath}
\usepackage{multirow}
\usepackage{caption}
\usepackage{subcaption}

\usepackage{graphicx}
\graphicspath{{./figures/}{./}}
\usepackage{threeparttable}

\graphicspath{{figures/}}

\begin{document}

\title{LightCurveLynx: Forward Modeling of Time-Domain Surveys with Application to ZTF SN Ia DR2
}

\author[0000-0002-5995-9692]{Mi~Dai}
\affiliation{
        Pittsburgh Particle Physics, Astrophysics, and Cosmology Center (PITT PACC),
        Department of Physics and Astronomy, University of Pittsburgh,
        Pittsburgh, PA 15260, USA
}
\email{mi.dai@pitt.edu}  

\author[0009-0009-2281-7031]{Jeremy~Kubica}
\affiliation{McWilliams Center for Cosmology and Astrophysics, Department of Physics, Carnegie Mellon University, Pittsburgh, PA 15213, USA}
\email{jkubica@andrew.cmu.edu}  

\author[0000-0001-7179-7406]{Konstantin~Malanchev}
\affiliation{McWilliams Center for Cosmology and Astrophysics, Department of Physics, Carnegie Mellon University, Pittsburgh, PA 15213, USA}
\email{malanchev@cmu.edu}  

\author[0000-0002-8676-1622]{Alex~I.~Malz}
\affiliation{Space Telescope Science Institute, Baltimore, MD 21218, USA}
\email{aimalz@nyu.edu}  

\author[0000-0001-5028-146X]{Olivia~Lynn}
\affiliation{McWilliams Center for Cosmology and Astrophysics, Department of Physics, Carnegie Mellon University, Pittsburgh, PA 15213, USA}
\email{olynn@andrew.cmu.edu}  

\author[0000-0001-5576-8189]{Andrew~Connolly}
\affiliation{DiRAC Institute and the Department of Astronomy, University of Washington, 3910 15th Ave NE, Seattle, WA 98195, USA}
\email{ajc@astro.washington.edu}  

\author[0000-0003-2271-1527]{Rachel~Mandelbaum}
\affiliation{McWilliams Center for Cosmology and Astrophysics, Department of Physics, Carnegie Mellon University, Pittsburgh, PA 15213, USA}
\email{rmandelb@andrew.cmu.edu}  

\author[0000-0001-7113-1233]{W.M.~Wood-Vasey}
\affiliation{
        Pittsburgh Particle Physics, Astrophysics, and Cosmology Center (PITT PACC),
        Department of Physics and Astronomy, University of Pittsburgh,
        Pittsburgh, PA 15260, USA
}
\email{wmwv@pitt.edu}  

\correspondingauthor{Mi~Dai}
\email{mi.dai@pitt.edu} 

\begin{abstract}

We present \texttt{LightCurveLynx}, a flexible and extensible software framework for end-to-end forward modeling time-domain light curves. 
Given the growing need for realistic simulations in the time-domain astronomy community, \texttt{LightCurveLynx} is designed to support a wide range of applications, including the development and validation of analysis pipelines, the optimization of survey strategies, and simulation-based inference studies. Realistic simulations can be generated from real survey metadata, forecasted survey plans, or user-defined mock survey strategies.
We demonstrate the functionality of \texttt{LightCurveLynx} by generating a realistic simulation of Type Ia supernovae that is representative of the ZTF SN Ia Data Release 2 dataset and perform extensive comparisons between the simulated and observed samples to validate the software. 
The simulation shows excellent agreement with the data in parameter distributions ({with the Kullback-Leibler divergence values around 0.01-0.02}) and in noise properties. The Hubble diagram generated from the simulation also indicates that the sample is complete up to redshift 0.06, which is consistent with previous studies.
Our results confirm that \texttt{LightCurveLynx} is robust, accurate, and ready for community use and contribution.

\end{abstract}


\newcommand{\hl}[1]{\textcolor{red}{#1}}

\section{Introduction}

Realistic simulations of light curves are an essential component of modern time-domain astrophysics. 
They enable the development and evaluation of classification algorithms \citep{Kessler2010, Kessler2019, Hlozek2023}, the optimization of survey strategies \citep{Lochner2018, Gris2024, Rose2025}, the validation of analysis pipelines \citep{Kessler2009}, the characterization and correction of systematic biases \citep{Kessler2017, Boyd2024, Vincenzi2024}, and the inference of astrophysical and cosmological parameters \citep{Weyant2013, Jennings2016}. 
While many groups in the astronomy community have independently created impressive simulation packages for their science domains, these solutions remain fragmented and features vary from solution to solution. At the same time, the demand for a unified, flexible, scalable, and user-friendly simulation framework has grown rapidly as new facilities with unprecedented discovery power and photometric precision, such as the Vera C. Rubin Observatory’s Legacy Survey of Space and Time \citep[LSST,][]{LSSTScienceBook2009,Ivezic2019} and the Nancy Grace Roman Space Telescope \citep{Spergel2015}, begin operations.

\lynx, developed by the LINCC Frameworks team, 
is designed to meet these community needs. 
It provides a unified, open-source framework for forward-simulation of transient and time-varying phenomena that supports both a variety of surveys and model types. 
It wraps a range of popular forward modeling packages, allowing users to combine and compare models, and includes numerous observational effects.
Supported by software development best practices, it facilitates sustainable user contributions with documentation, unit tests, and continuous integration.

In this paper, we present \lynx\ in the context of a use case in which we simulate Type Ia supernova (SN Ia) light curves observed by the Zwicky Transient Facility (ZTF, \citealt{Bellm2019}). 
We perform extensive comparisons against real data from the ZTF SN Ia Data Release 2 sample (hereafter \ztfsn, \citealt{Rigault2025}) demonstrating both the functionality and fidelity of \lynx.

We provide an overview of the \lynx\ software framework in Section \ref{sec:lynx_descrip}. 
In Section \ref{sec:sim_ztf}, we describe in detail the inputs and methodology required to generate realistic simulations of the \ztfsn\ sample, including survey characteristics, noise modeling, supernova and host-galaxy models, parameter generation, and selection effects. 
In Section \ref{sec:results}, we present quantitative comparisons between the simulated and observed \ztfsn\ samples. 
We discuss our results and conclude in Section \ref{sec:discussion}.

\section{\lynx}\label{sec:lynx_descrip}

\begin{figure}[h]
    \centering
    \includegraphics[width=0.8\linewidth]{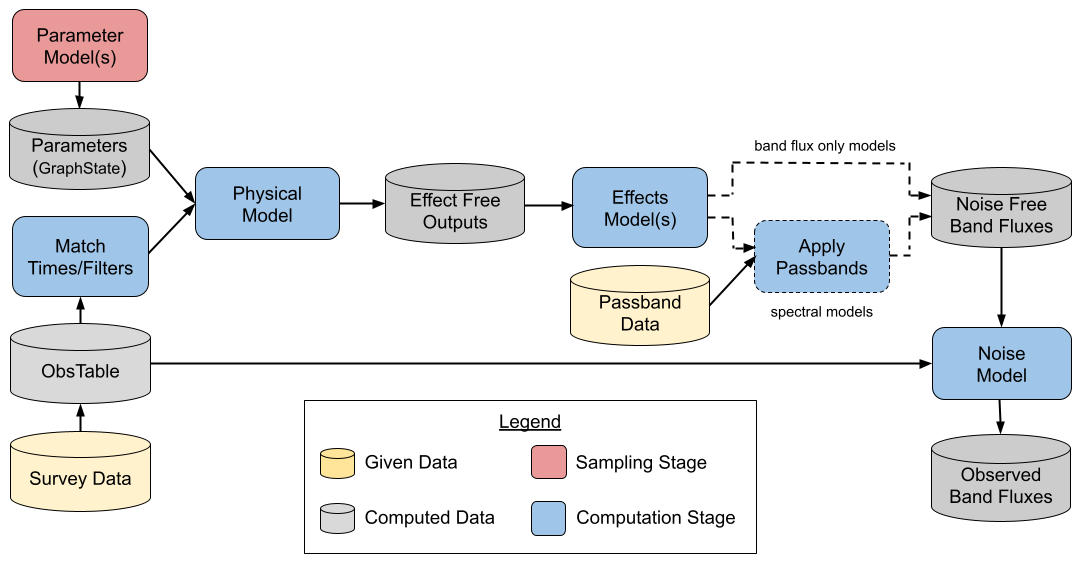}
    \caption{Schematic overview of the \lynx\ framework, showing the major components and data flow in a typical simulation pipeline.}
    \label{fig:lynx-intro}
\end{figure}

\lynx\ is a Python-based forward-modeling framework for generating realistic time-domain light curve simulations, including both transients and variable sources. 
The software is designed to be flexible, scalable, and user-friendly. 
Its modular architecture allows users to integrate \lynx\ into existing pipelines or construct entirely new simulation workflows by replacing or extending individual components. 
Figure~\ref{fig:lynx-intro} illustrates the main components of the \lynx\ framework.

A typical \lynx\ simulation begins with the definition of \textit{Parameter Models}, which specify the statistical distributions and relationships among all physical parameters (the priors). 
These dependencies can be represented as a directed acyclic graph (DAG), where parameter distributions can themselves depend on the values of hyper-parameters.
\lynx\ provides built-in functionality for DAG construction, manipulation, and efficient sampling. 
\lynx\ generates a full ensemble of parameters (stored as a \textit{GraphState} object) that are fed into the \textit{Physical Model}.

An essential element of the simulation pipeline is the \textit{Physical Model} modules, which provide a recipe for generating samples from a time-varying flux density (at either the spectral or photometric level) of an object with given parameters. 
In addition to user-implemented physical models, \lynx\ wraps a variety of popular forward modeling packages, such as \sncosmo\ \citep{Barbary2016,barbary_2025_15019859}, \pzflow\ \citep{Crenshaw2024} and \redback\ \citep{sarin_2024}, allowing users to use a range of models within a consistent framework. 

A list of \lynx-wrapped packages can be found on \lynx's ReadtheDocs page\footnote{\url{https://lightcurvelynx.readthedocs.io/en/latest/}}, which also includes extensive API documentation and usage tutorials.

A physical model is combined with an \textit{ObsTable}, which provides the observational characteristics of a survey, such as cadence, zeropoint, sky noise, and seeing conditions, to determine the observation times and filters for each object. 
\lynx\ greatly reduces computational cost and storage requirements by only sampling the flux density at times when the survey actually observes the object. 
The output flux densities are then generated from the physical model and its parameters at each such time. 
At this stage, the model fluxes are in the rest frame and noise-free.

Physical effects such as host-galaxy extinction or gravitational lensing can be defined through \textit{Effect Model} modules that are applied to the simulated fluxes to generate realistic observational conditions. 
The modeled flux density is then transformed to the observer frame. If the simulation has been done at the spectral level, the flux densities are integrated through the relevant \textit{Passbands} to produce broadband fluxes. 
Finally, observational noise is added to the fluxes using a \textit{Noise Model} that depends on both the flux level and the  \textit{ObsTable} information, yielding simulated measurements that accurately reflect survey conditions.

A full description about \lynx's design and functionalities is presented in \cite{lynx_software}.

\section{Simulating the ZTF SN Ia Data Release 2 sample}\label{sec:sim_ztf}

\cite{Rigault2025} presented \ztfsn, a data set of 3628 SNe Ia discovered by the Zwicky Transient Facility \citep[ZTF,][]{Bellm2019} from 2018 to 2020. 
The SNe are spectroscopically classified, with redshifts ranging from 0.002 to 0.29. 
The data release includes forced photometry light curves at the SN locations, SN metadata, and observing logs providing per-exposure information for simulating cadence and realistic noise conditions.
The host galaxy data, such as angular coordinates, redshift, local and global stellar mass, and $g-r$ color, are also included in the data release. 

We describe the necessary components in the context of generating the \ztfsn\ simulation, following the workflow described in Figure \ref{fig:lynx-intro}. Users can replace these components individually to generate customized simulations for their specific science cases. All the data and analysis notebooks are available in a dedicated GitHub Repository\footnote{\url{https://github.com/mi-dai/lightcurvelynx_ztf_sims}.\label{fn:repo}}.

\subsection{Parameter Models and Physical Models}

We describe the Physical Models, including the host-galaxy model and the SN Ia Spectral Energy Distribution (SED) model, as well as the Parameter Models that are used to generate the SN Ia populations, including data-driven parameter models for SN and host galaxy parameters, cosmological parameters, and the SN Ia rate.

\subsubsection{Host-galaxy Model}

We generate a simple host galaxy model, with only RA, Dec, and host stellar mass as parameters. 
The SN angular separation from the host galaxy is generated using a distribution of physical separation roughly following that in Figure 3 of \cite{Gupta2016}, an exponential distribution, $f(x;1/\beta) = 1/\beta \times \exp{(-x/\beta)}$, 
with the exponential scale $\beta$ set to 5 kpc. 
We relate the host galaxy stellar mass to the SN light curve parameters using a data-driven approach described in Section~\ref{subsubsec:pzflow}. 
\lynx\ can be extended to support more complex simulations of host galaxy and SN relations, for example, SN locations based on the galaxy light profile or SN parameters correlated with additional host properties (e.g., star formation rate and metallicity; \citealt{Gagliano2021,Lokken2023,elasticc}).

\subsubsection{SN Ia SED Model}

We use the \sncosmo\ implementation of the SALT3 model \citep{Kenworthy2021}, which is a newer version of the SALT model \citep{Guy2007, Guy2010}, because the SALT SED model is commonly used in SN Ia cosmology analyses. 
The flux density is defined as:
\begin{equation}
F(t, \lambda) = x_0 \cdot \left[ M_0(t, \lambda) + x_1 \cdot M_1(t, \lambda) \right] \cdot \exp\left( c \cdot CL(\lambda) \right) ,
\end{equation}
where $M_0$ is the average spectral sequence of an SN Ia, $M_1$ is the first-order variation, and $CL$ is the color variation law. 
$x_0, x_1, c$ are parameters that describe individual supernova properties and can be fitted using observed SN Ia light curves. $x_0$ describes the overall normalization, which is related to the SN peak magnitude; $x_1$ encodes the shape of the light curve; $c$ describes the SN color.

The SALT3 model is only defined for rest-frame phases between -20 days and 50 days, and for rest-frame wavelengths between 2000~\AA\ and 11000~\AA. Extrapolations are applied to the original SED model beyond the model range in both the time and wavelength dimensions. Since the signals are usually low before -20 days, we pad zeros to the SED before -20 days. We apply a linear decay in the magnitude space with decay rate at 0.02 mag per day for phases after 50 days. For wavelengths, we use zero padding on both sides of the SED, since those wavelength regions are usually dominated by noise.

\subsubsection{Data-driven Parameter Models with \texttt{pzflow}}\label{subsubsec:pzflow}

\pzflow\ \citep{Crenshaw2024} is a python package that can model a potentially high-dimensional joint or conditional distribution of data using normalizing flows. 
We used \pzflow\ to train on the \ztfsn\ data including the SN parameters and the host galaxy's stellar mass. Selection functions of the SN parameters are also calculated and accounted for using \pzflow. We describe the selection effects and the methodology for accounting for selection effects in Section \ref{subsec:selection_effects}.

\subsubsection{Rate Model and Total Number of SNe}

Following \cite{Amenouche2025}, we use a volumetric rate of $R = 2.35 \times 10^{-5} \text{Mpc}^{-3} \text{year}^{-1}$, derived from the ZTF Bright Transient Survey data \citep{Perley2020}. 
As discussed in \cite{Amenouche2025}, the rate is consistent with other studies \citep{Dilday2010,Frohmaier2019} for the same redshift range. 

The total number of SNe Ia to generate is calculated as $N_{\rm total}=f_{\rm corr}\Omega T \int R \, dV(z) dz / (1+z)$, where $dV(z)$ is the differential comoving volume as a function of redshift, $\Omega$ is the survey volume in solid angle, T is the survey length in years in observer frame. We also apply a correction factor, $f_{\rm corr}=0.867$. This correction factor corrects for potential missing detections due to CCD gaps \citep{Dekany2020}.

\subsubsection{Cosmology and Nuisance Parameters}

We assume a flat $\Lambda$CDM cosmology with $\Omega_\text{m} = 0.315$ and $H_0 =  70 \, \mathrm{km\,s^{-1}\,Mpc^{-1}}$. 
Using the SALT parameters, the distance modulus of SN Ia is defined as below following the Tripp Relation \citep{Tripp1998}. 

\begin{equation}\label{eq:tripp}
    \mu_\text{SN} = m_B + \alpha x_1 - \beta c - M_B
\end{equation}

where $m_B = -2.5 \log_{10} x_0 + 10.635$, which roughly represents the SN peak magnitude in the B band. 
$\alpha$, $\beta$ encode the amount of linear correction for SN Ia standardization. 
$M_B$ is the B band absolute magnitude for the SN Ia. 

Following \cite{Rigault2025}, we use $\alpha = 0.16$ and $\beta = 3.05$, which is derived by fitting the Hubble Diagram of a volume-limited sample from \ztfsn\ \citep{Ginolin2025}.  
We assume the B-band absolute magnitude of $M_B = -19.08$ with a Gaussian scatter of $0.17$ mag. 
To determine the mean of the $M_B$ distribution, we take the data from \cite{Rigault2025}, correct for the shape and color relation using Eq. \ref{eq:tripp}, and fit a Gaussian model for the mean and standard deviation. 
Our fit returns $\text{Mean}=-19.11$ and $\text{std}=0.17$. 

After running a full simulation and applying all selection cuts, we noticed a shift of $-0.03$ in the Gaussian Mean value. 
We apply this shift to account for the selection effects in $M_B$ (thus $\textrm{Mean}(M_B) = -19.08$). 
We do not include a host galaxy mass dependency as done by other analyses, as the formulation thereof is beyond the scope of this paper. 
However, we note that such a dependency can also be modeled in \lynx\ using the data-driven approach described in Section \ref{subsubsec:pzflow}.

\subsection{Observing Logs}\label{subsec:obslog}

We combine the observing log from \cite{Rigault2025} with the ZTF metadata database (DB) from the ZTF data release note\footnote{\url{https://irsa.ipac.caltech.edu/data/ZTF/docs/releases/dr23/ztf_release_notes_dr23.pdf}} to obtain all necessary information for generating our simulations. The observing log contains CCD visit information from June 2018 to February 2021\footnote{However, the paper states that the logs are from March 2018 to December 2020.}, including observation times, coordinates, filter names, zero points, and 5$\sigma$ limiting magnitudes. The metadata DB supplements this with information required for realistic noise calculations, primarily the point-spread function.
The observing log is provided per CCD quadrants for each exposure. 
For simplicity, we take the medians of the zero points and the 5$\sigma$ magnitude limits of all CCD quadrants in each exposure and drop the duplicated rows. 
The standard deviation of the zero points is around 0.1 mag. 
The sky background for each exposure is needed for estimating the noise. However, we find the sky background values provided in the metadata DB do not appear to represent the actual sky background measurement in the images. We therefore derive the sky background from existing information using Eq.~\eqref{eq:noise}, assuming that the 5$\sigma$ magnitude limit is the magnitude at which $\texttt{flux}/\texttt{flux error}= 5$. We describe the caveats in deriving the sky background in Appendix~\ref{appx:skybg}.

The columns that are necessary for generating the simulation are listed in Table~\ref{obslog}. 
The combined observing log is available from the GitHub repository\footref{fn:repo}.

\begin{table}[h!]
\begin{threeparttable}
\caption{Observing log column descriptions}
\label{obslog}
\begin{tabular}{|p{2.5cm}|p{3.5cm}|p{8cm}|}
\hline
\textbf{Observing Log Column Name} & \textbf{Source} & \textbf{Description} \\ \hline
\texttt{mjd}        & \cite{Rigault2025} & MJD values of the observations. \\ \hline
\texttt{filter}     & \cite{Rigault2025} & Filter names of the observations. Original column name: \texttt{band}. \\ \hline
\texttt{ra}         & \cite{Rigault2025} & Right Ascension of the observations. Original column name: \texttt{fieldra}. \\ \hline
\texttt{dec}        & \cite{Rigault2025} & Declination of the observations. Original column name: \texttt{fielddec}. \\ \hline
\texttt{maglimit}   & \cite{Rigault2025} & 5$\sigma$ magnitude limit of the observations. \\ \hline
\texttt{zp\_abmag}  & \cite{Rigault2025} & Median zero point across all CCD quadrants (magnitude corresponding to 1~ADU). Original column name: \texttt{zp}. \\ \hline
\texttt{zp\_nJy}    & Derived & Zero point (nJy/e$^-$). Converted from \texttt{zp\_abmag}. \\ \hline
\texttt{fwhm}       & Metadata DB & Full width at half maximum of the PSF (pixel). \\ \hline
\texttt{maglim}     & Metadata DB & 5$\sigma$ magnitude limit of the observations. (This is different from \citealt{Rigault2025}. See Appendix \ref{appx:skybg}.) \\ \hline
\texttt{infobits}   & \cite{Rigault2025} & Information bits of the observations.\tnote{a} \\ \hline
\texttt{sky\_adu\_ztfsn} & Derived & Sky background (ADU). Derived as described in Section~\ref{subsec:obslog} using \texttt{maglimit}. \\ \hline
\end{tabular}
\begin{tablenotes}
\item[a] See the ZTF Science Data System (ZSDS) Explanatory Supplement: \url{https://irsa.ipac.caltech.edu/data/ZTF/docs/ztf_explanatory_supplement.pdf}.
\end{tablenotes}
\end{threeparttable}
\end{table}

\subsection{Time and Filter Matching}
\lynx\ generates a list of times and filters for computing the Physical model by matching a given sky position to the observing log. The matching is determined by computing whether the random sky position falls in the camera footprint, given the CCD configuration. For this analysis, we used the camera field dimensions listed in Table 1 of \cite{Dekany2020}. The camera field dimensions represent the full camera footprint, including CCD gaps. Since we do not model CCD gaps in the input camera footprint for our simulation, our simulation include observations that are missed in the actual survey due to these gaps. The fill factor is $86.7 \%$ \citep{Dekany2020}. We use this number to roughly approximate the number of SNe that may be reduced in the sample due to CCD gaps. We note that the actual loss number is likely different. A full simulation with the exact CCD configurations is necessary to recover the actual factor. \lynx\ supports simulations using exact CCD layouts; however, we use the approximations described above as a tradeoff for computational efficiency.

\subsection{Effects}

\lynx\ supports built-in and user-defined physical or observational effects. For this paper, we include the Milky Way extinction as an effect. We obtain the extinction parameter E(B-V) from \cite{Schlegel1998, Schlafly2011} using the \texttt{sfdmap2} package \citep{sfdmap2} wrapped by \lynx, and apply extinction to the observer-frame SN SED, before integrating over the passbands to generate broadband photometry. We use the extinction curve from \cite{Fitzpatrick1999} as provided in the \texttt{dust\_extinction} package \citep{Gordon2024}.

\subsection{Filter Transmission Curves}

The \ztfsn\ data include three passbands, g, r, and i. 
We obtain the filter transmission curves from \sncosmo\footnote{\url{https://sncosmo.readthedocs.io/en/stable/bandpass-list.html}}. \lynx\ provide a variety of ways to obtain transmission curves from different sources.

\subsection{Noise Estimation}\label{subsec:noise}

The total noise standard deviation in electrons for a point source is calculated following standard photon noise calculation assuming Gaussian limit for the Poisson electron count distribution:

\begin{align}\label{eq:noise}
\sigma_{\text{total}}^2
&= \sigma_{\text{signal}}^2 + \sigma_{\text{sky}}^2
   + \sigma_{\text{dark}}^2 + \sigma_{\text{read}}^2
   + \sigma_{\text{zp}}^2\\[6pt]
&= S + \text{Sky} \, N_{\text{eff}}
   + D \, t \, N_{\text{eff}}
   + \sigma_{\text{read}}^2 \, N_{\text{eff}}
   + \left( S \, \frac{\sigma_\texttt{zp\_mag}}{\texttt{zp\_mag}} \, \frac{\ln{10}}{2.5} \right)^2 \, .
\end{align}

\begin{equation*}
\begin{aligned}
\text{where} \quad 
& \sigma_{\text{total}} &&= \text{total noise (in electrons)} \\
& S &&= \text{signal electrons (shot noise is } \sqrt{S} \text{ electrons)} \\
& \text{Sky} &&= \text{sky background count (in electrons/pixel}^2)\\
& D &&= \text{dark current (in electrons/pixel}^2\text{/second}) \\
& t &&= \text{exposure time (in seconds)} \\
& \sigma_{\text{read}} &&= \text{readout noise (in electrons/pixel}) \\
& N_{\text{eff}} &&= \text{PSF Effective Area (pixel}^2)\\
& \texttt{zp\_mag} && = \text{Zeropoint (magnitude corresponding to 1 electron)}\\
& \sigma_\texttt{zp\_mag} && = \text{Zeropoint error (mag)}
\end{aligned}    
\end{equation*}

For ZTF, the CCD characterizations (i.e. $D$, $\sigma_{\text{read}}$) are obtained from Table 3 in \cite{Dekany2020}. 
Some values are converted to electrons using $\text{CCD gain} = 6.2 e^-$. 
$N_\text{eff}$ is calculated from the PSF's Full width at half maximum (FWHM) assuming Gaussian PSF. 
For our simulations generated, we set $\sigma_\texttt{zp\_mag} = 0$.

\subsection{Selection Effects}\label{subsec:selection_effects}

After the simulation is generated, we apply selection effects to obtain a sample consistent with the data selection criteria.
We included detection cuts, spectroscopic follow-up efficiency, and light curve quality cuts.

    \paragraph{Detection}
    
    We define detection as an observation with signal-to-noise-ratio (SNR) greater than 5 and require at least one detection for the simulated light curve to be included in the final sample.

     \paragraph{Spectroscopic efficiency}
    
    Since \ztfsn\ is a spectroscopically confirmed sample, we apply the same spectroscopic efficiency as used in \cite{Rigault2025}. 
    The spectroscopic efficiency curve is modeled as a survival sigmoid
    \begin{equation}
        1 - \mathcal{S}(m; m_0, s) = 1 - \left( 1 + e^{s (m - m_0)} \right)^{-1} ,
    \end{equation}
    where $m_0 = 18.8, \; s = 4.5$. 
    This curve matches the spectral completeness curve for the ZTF Bright Transient Survey (BTS, \citealt{Perley2020}).
   
     \paragraph{Light curve quality cuts}
    
    We apply the same light curve quality cuts following \cite{Rigault2025}, which requires detections at 7 different phases, at least 2 detections before and 2 detections after the peak magnitude, and detections in at least two different bands within the phase range of [-10, 40] days. 
    However, for simplicity, we do not fit the light curve to determine the time of peak, instead using the time of the maximum flux as an approximation. 

\paragraph{Modeling selection effects in parameter models}

In order to account for selection effect in the input parameter models for the SALT parameters $x_1$ and c, and the host galaxy stellar mass, the following steps are performed. 

\begin{enumerate}
    \item Generate simulations with flat distributions of the SN parameters $x_1$ and $c$.
    \item Apply selection cuts as described in Section~\ref{subsec:selection_effects}.
    \item Compute and fit for the selection functions for $x_1$ and $c$ separately. 
    We fit the discrete selection function with an exponential function, $y = a e^{b x} +c$, to obtain a smooth selection function. 
    For $x_1$, we require $a > 0$ and $b > 0$ so the selection function follows the broader-brighter relation (more objects with larger $x_1$ values will be selected due to their brightness being higher; \citealt{Phillips1993, Tripp1998}). 
    For $c$, we require $a > 0$ and $b < 0$, following the bluer-brighter relation (more objects with smaller $c$ values will be selected due to their brightness being higher; \citealt{Riess1996, Tripp1998}).
    \item Train a \pzflow\ model with existing $x_1$, $c$ data to obtain distributions after selection.
    \item Inverse-apply selection function to obtain the $x_1$, $c$ distribution before selection. 
    (Note that the SALT3 parameters $x_1$ and $c$ are orthogonal by design, so we apply selection functions separately and ignore the covariance between them. 
    Alternatively, a joint selection function for a set of parameters can be estimated using similar approaches.)
    \item Draw samples from the new $x_1$, $c$ distribution and train another \pzflow\ model on host mass conditioned on $x_1$, $c$.
\end{enumerate}

\begin{figure*}
    \centering
    \begin{subfigure}{0.48\linewidth}
        \centering
        \includegraphics[width=\linewidth]{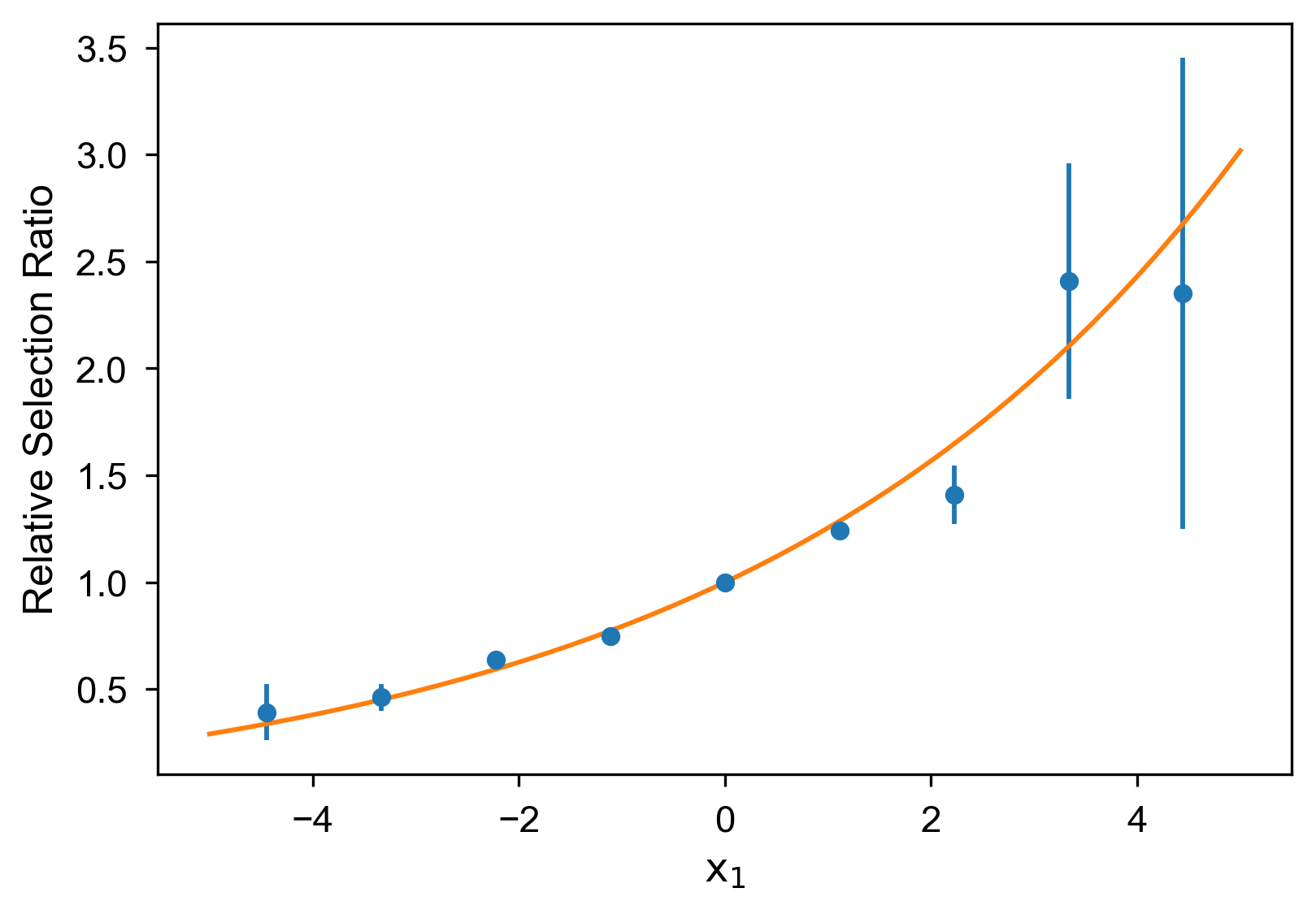}
        \label{fig:x1_selection}
    \end{subfigure}
    \hfill
    \begin{subfigure}{0.48\linewidth}
        \centering
        \includegraphics[width=\linewidth]{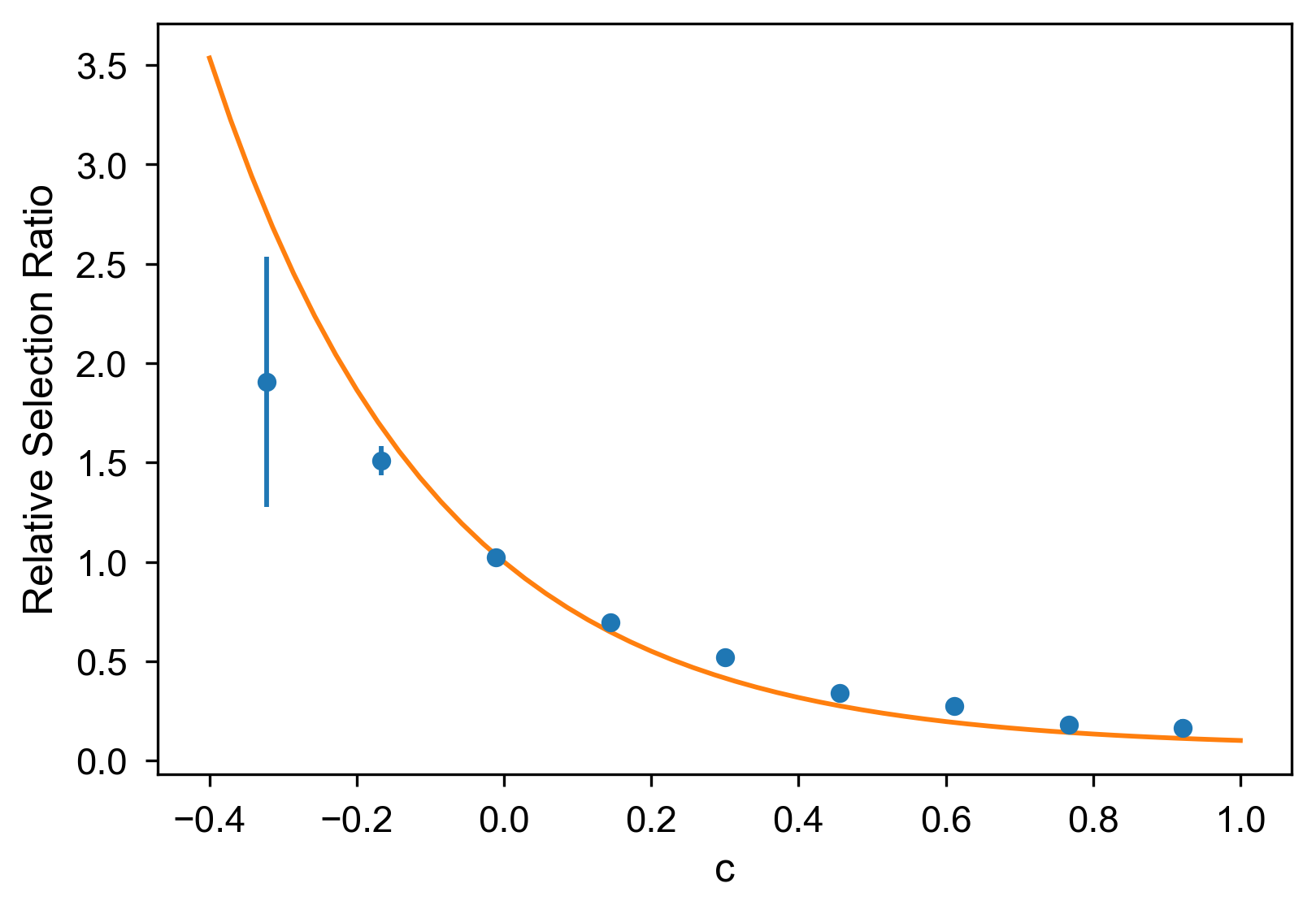}
        \label{fig:c_selection}
    \end{subfigure}
    \caption{The selection functions for the SN Ia light curve parameters $x_1$ and $c$. 
    Solid orange lines represent the selection functions derived following the procedure in Section \ref{subsubsec:pzflow};
    blue dots represent the recovered selection function from the simulations. 
    The selection functions are re-scaled so that the relative selection ratio is 1 when $x_1$ and $c$ are zero.}
    \label{fig:selection_function}
\end{figure*}

These steps are demonstrated in a publicly available Jupyter notebook\footnote{\url{https://github.com/mi-dai/lightcurvelynx_ztf_sims/blob/main/train_pzflow.ipynb}}.
The derived selection functions for $x_1$ and $c$, together with the recalculated ones using simulation, are shown in Figure \ref{fig:selection_function}.

\subsection{Light Curve Fitting}

For validation purposes, we fit the simulated light curves using \sncosmo\ and compare the fitted values to the simulated values. 
The light curve fitting is only performed on the sample after light curve quality cuts. The same SALT3 model for simulation is used for light curve fitting. We limit the rest frame phase range [-10, 40] days, following \cite{Rigault2025b}. The bounds for $x_1$ and $c$ parameters are set as [-5, 5] and [-0.4, 1], respectively. The best-fit parameters all agree with the simulated parameters within the uncertainties.

\section{Results}\label{sec:results}

We present our results in three areas: comparisons between the simulation and data, the recovered Hubble diagram, and simulation performance.

\subsection{Data and Simulation Comparisons}
We show comparisons between the data and simulation in terms of individual light curves, sample statistics, parameter distributions, and noise properties.

\subsubsection{An Example Light Curve}

Before comparing the full sample, we first simulate a random SN in the \ztfsn\ sample using its location and best-fit SALT parameters, and then compare with the data. 
We query the observing log and match the exact CCD quadrant as listed in the data to avoid introducing any discrepancy due to variations in the CCDs.  

\begin{figure}
    \centering
    \includegraphics[width=0.5\linewidth]{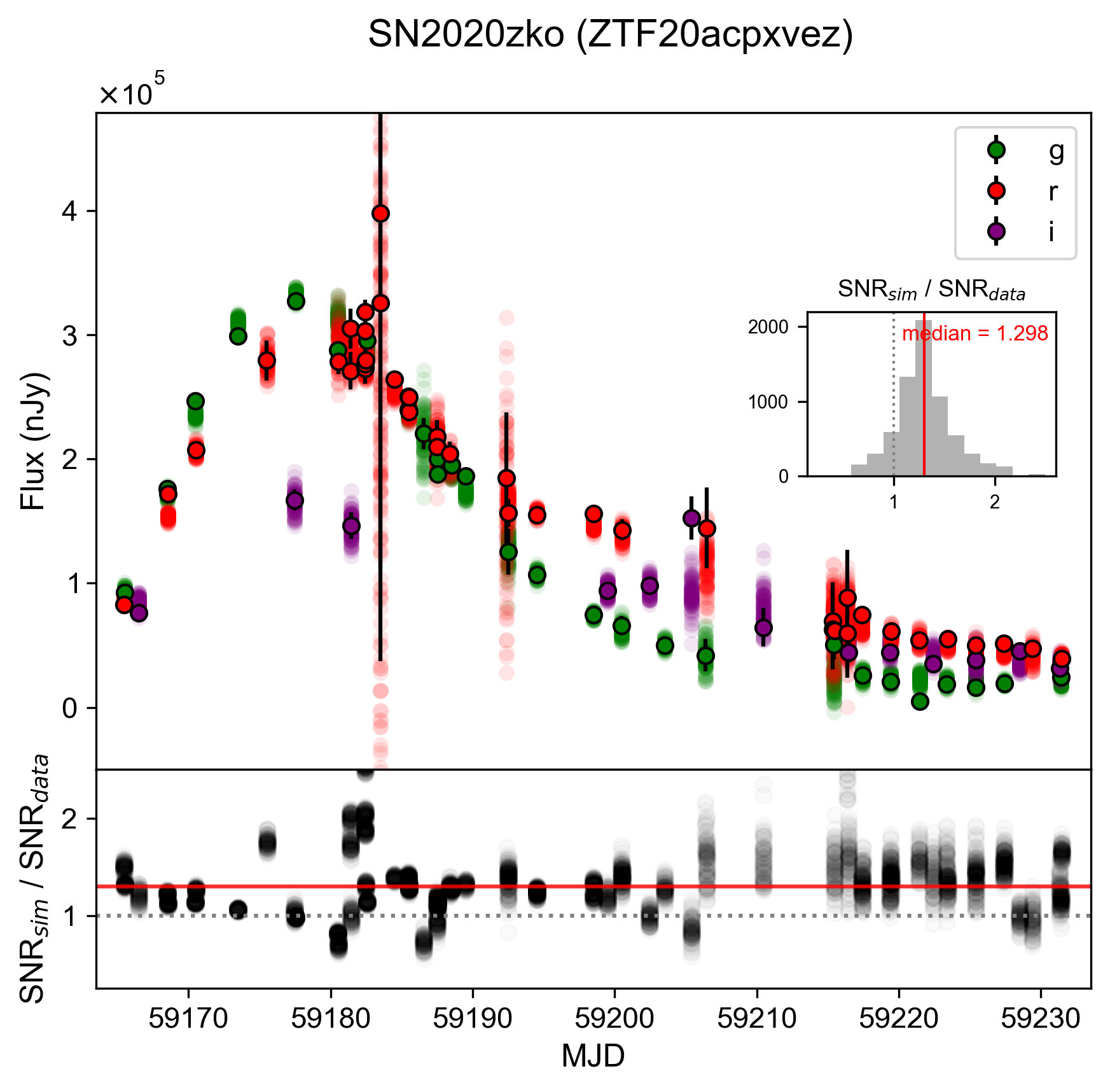}
    \caption{Top panel: 
    An example of a simulated light curve of a randomly selected real \ztfsn\ object.
    The solid dots are the \ztfsn\ data for this object, the transparent points show 100 realizations of the simulated light curves using the same set of best-fit light curve parameters for this object. 
    The spread in the points represents random realizations of the simulated band fluxes from the same Gaussian errors. 
    Bottom panel: 
    The ratio of the SNR between the simulations and the \ztfsn\ data.}
    \label{fig:example_lc}
\end{figure}

Figure \ref{fig:example_lc} shows a comparison between a real \ztfsn\ light curve (SN2020zko) and 100 realizations of simulated light curves using the same set of light curve parameters. 
The simulated flux values appear consistent with the data, with the median value of the simulation to data flux ratios being 0.989. 
For this example SN, the median values of the simulation to data flux error ratios are 0.766, and the median values of the simulation to data SNR ratios, as shown in Figure~\ref{fig:example_lc}, is 1.295. 
The values indicate that the flux errors of this SN is underestimated by $\sim 23\%$. 
This is consistent with our assumptions, since we infer the sky background from the published 5$\sigma$ limiting magnitude from the science images (see Appendix \ref{appx:skybg}). 
While actual light curves are obtained by performing forced photometry on difference images, and we do not include the error contribution from the template images, the underestimation in the flux errors is expected. 
In a similar effort, \cite{Amenouche2025} also found that the flux errors are underestimated when estimating the sky noise using the magnitude limit from the science image, and applied a per-band correction factor to the simulated flux errors to resolve the discrepancies between simulation and data. 
The correction factors are 1.23, 1.17 and 1.20 for g, r and i bands respectively, which is consistent with the level of underestimation we see in this SN. 
We further simulated the full \ztfsn\ sample and found the median of the simulation to data SNR ratio ($\textrm{SNR}_\textrm{sim}/\textrm{SNR}_\textrm{data}$) is 1.12. 
Future investigations will be needed to understand the exact source of the error underestimation, which is beyond the scope of this paper. 
While the exact precision requirements for simulations are likely dependent on the specific science cases, \lynx\ can be extended to include more complex error simulations (e.g. adding the template contributions from the difference image analyses, or modeling host galaxy residuals from subtraction). 
A correction factor as in \cite{Amenouche2025} can also be applied using \lynx\ to resolve the simulation and data discrepancy if the source of the discrepancy is beyond the scope of the analysis.

\subsubsection{Sample Statistics}

We show in Table \ref{tab:sn_cuts} the total number of SNe simulated and passed each selection criteria, with a comparison to \cite{Rigault2025}. 
The total sample size in our simulation is about $20\%$ larger compared to the \ztfsn\ data. 
This may be due to the uncertainty in the volumetric rate estimation, or the slightly underestimated flux errors which will make more SNe passing the selection cuts, or a slightly different total survey area (we used the observing log to estimate the total sky coverage, however, the observing log may include fields that are not part of the science area). 
We do not include host galaxy light profile or dust properties in the simulation, which may also result in a larger sample number, since some of the SNe may not be detectable due to these effects. The percentages passing each cut are similar between the simulation and data.

We further compare sample statistics on number of observations, number of detections, and median cadences of each filter in Table \ref{tab:sim_data_summary}. 
For the simulation, we compute the numbers using the sample that pass the spectroscopic selection cuts. 
This sample should have similar statistics with the full \ztfsn\ data. 
The observing log from \cite{Rigault2025} includes observations from June 19, 2018 to Feb 28, 2021, while the actual light curves appear to have additional data points outside of this range. 
For comparison, we cut the \ztfsn\ data to include the same range as the observing log covers. 
We also apply cuts to exclude observations that have unphysical errors, are extreme outliers, have a cloudy flag, or have infobits $>0$ based on the \texttt{flag} column in the \ztfsn\ data.
Our results show that all listed metrics are comparable between the simulation and data, while the simulation generally has slightly larger numbers of observations and detections per object, possibly due to our neglect of CCD gaps.
We calculate the median cadence by taking the difference of the MJDs in each object, finding the median cadence of each object, and report the median value of the median cadences. 
We do the same for the subset that contains only detections. 
The median cadences for the g and r bands are nearly identical between simulation and data, while those for the i band show slightly larger differences. The i-band discrepancy may be an artifact of how the median cadence is calculated, as cadences are clustered in discrete numbers of days.
To separate out the effect that our light curve extrapolation may not be accurate enough to produce the same statistics for epochs out of the defined SALT range, we also show the number of detections within the rest frame phase range (-20, 50) days. 
The resulting numbers are closer for the two samples, but the \ztfsn\ sample have a slightly larger number of detections (in g and r) within the selected phase range.

\begin{table*}
\centering
\begin{tabular}{l|cc|cc}
\hline
\multirow{2}{*}{\textbf{Cuts}} 
    & \multicolumn{2}{c|}{\textbf{Simulation}} 
    & \multicolumn{2}{c}{\textbf{ZTFSNDR2}} \\
\cline{2-5}
 & \textbf{Number} & \textbf{\%} 
 & \textbf{Number} & \textbf{\%} \\
\hline
All simulated                        & 84{,}687  & 100.0               & --      & --                \\
After applying detection             & 43{,}303  & 51.13               & --      & --                \\
After spectroscopic selection        & 4{,}313   & 5.09 (100.0)        & 3{,}628 & 100.0 (100.0)     \\
After quality cuts                   & 3{,}346   & 3.95 (77.6)         & 2{,}960 & 81.6 (81.6)       \\
$x_1 >-3$ \& $x_1<3$                  & 3{,}224   & 3.81 (74.8)         & 2{,}899 & 79.9 (79.9)       \\
c $>-0.2$ \& c $<0.8$                & 3{,}159   & 3.73 (73.2)         & 2{,}861 & 78.9 (78.9)       \\
$\sigma_{t_0} < 1$                   & 3{,}062   & 3.62 (71.0)         & 2{,}836 & 78.2 (78.2)       \\
$\sigma_{x_1} < 1$                   & 3{,}031   & 3.58 (70.3)         & 2{,}822 & 77.8 (77.8)       \\
$\sigma_{c} < 0.1$                   & 3{,}010   & 3.55 (69.8)         & 2{,}809 & 77.4 (77.4)       \\
fitprob $>10^{-7}$                   & 2{,}992   & 3.53 (69.4)         & 2{,}667 & 73.5 (73.5)       \\
\hline
\end{tabular}
\caption{Comparison of the simulation and \ztfsn\ after successive selection cuts. 
Percentages in parentheses are relative to the spectroscopic-selected sample.}
\label{tab:sn_cuts}
\end{table*}

\begin{table*}
\centering
\renewcommand{\arraystretch}{1.25}
\begin{tabular}{
    >{\centering\arraybackslash}m{0.8cm}
    >{\centering\arraybackslash}m{1.5cm}
    >{\centering\arraybackslash}m{1.0cm}
    >{\centering\arraybackslash}m{1.25cm}
    >{\centering\arraybackslash}m{1.25cm}
    >{\centering\arraybackslash}m{1.4cm}
    >{\centering\arraybackslash}m{1.0cm}
    >{\centering\arraybackslash}m{1.0cm}
    >{\centering\arraybackslash}m{1.4cm}
    >{\centering\arraybackslash}m{1.35cm}
    >{\centering\arraybackslash}m{1.35cm}
}
\hline
\textbf{Filter} &
\textbf{Sample} &
$\mathbf{N_{event}}$ &
$\mathbf{N_{obs}}$ &
$\mathbf{N_{det}}$ &
$\mathbf{N_{det,win}}$ &
$\mathbf{\langle N_{obs}\rangle}$ &
$\mathbf{\langle N_{det}\rangle}$ &
$\mathbf{\langle N_{det,win}\rangle}$ &
\textbf{Median cadence} &
\textbf{Median cadence det.} \\
\hline
\multirow{2}{*}{\textbf{g}}
 & Simulation  & 4313 & 1,588,686 & 115,106 & 94,609 & 368.35 & 26.69 & 21.94 & 1.03 & 2.00 \\
 & ZTFSNDR2 & 3592 & 1,267,460 & 102,326 & 84,162 & 352.86 & 28.49 & 23.43 & 1.03 & 1.99 \\
\hline
\multirow{2}{*}{\textbf{r}}
 & Simulation  & 4313 & 2,159,803 & 213,619 & 152,888 & 500.77 & 49.53 & 35.45 & 0.95 & 1.04 \\
 & ZTFSNDR2 & 3592 & 1,649,064 & 162,671 & 132,487 & 459.09 & 45.29 & 36.88 & 0.96 & 1.08 \\
\hline
\multirow{2}{*}{\textbf{i}}
 & Simulation  & 4313 & 306,993 & 26,289 & 19,800 & 71.18 & 6.10 & 4.59 & 3.09 & 4.00 \\
 & ZTFSNDR2 & 3592 & 224,896 & 17,786 & 15,558 & 62.61 & 4.95 & 4.33 & 3.99 & 4.57 \\
\hline
\end{tabular}
\caption{Sample statistics for the simulation and \ztfsn.}
\label{tab:sim_data_summary}
\end{table*}

\subsubsection{Parameter Distributions}

In this section, we compare distributions of several simulated parameters to the \ztfsn\ sample. 
The full \ztfsn\ sample is presented before the light curve quality cuts, and includes both detections and forced photometry of the detected objects. 
For the simulation, we apply detection criteria, spectroscopic efficiency, and light curve quality cuts (as described in \ref{subsec:selection_effects}), with flags saved to filter out or include them in the comparison accordingly. 
The ZTF sample also includes flags that can filter out objects that do not pass light curve quality or SALT parameter cuts.

    \paragraph{Redshift distribution}

\begin{figure}
        \centering
        \includegraphics[width=0.5\linewidth]{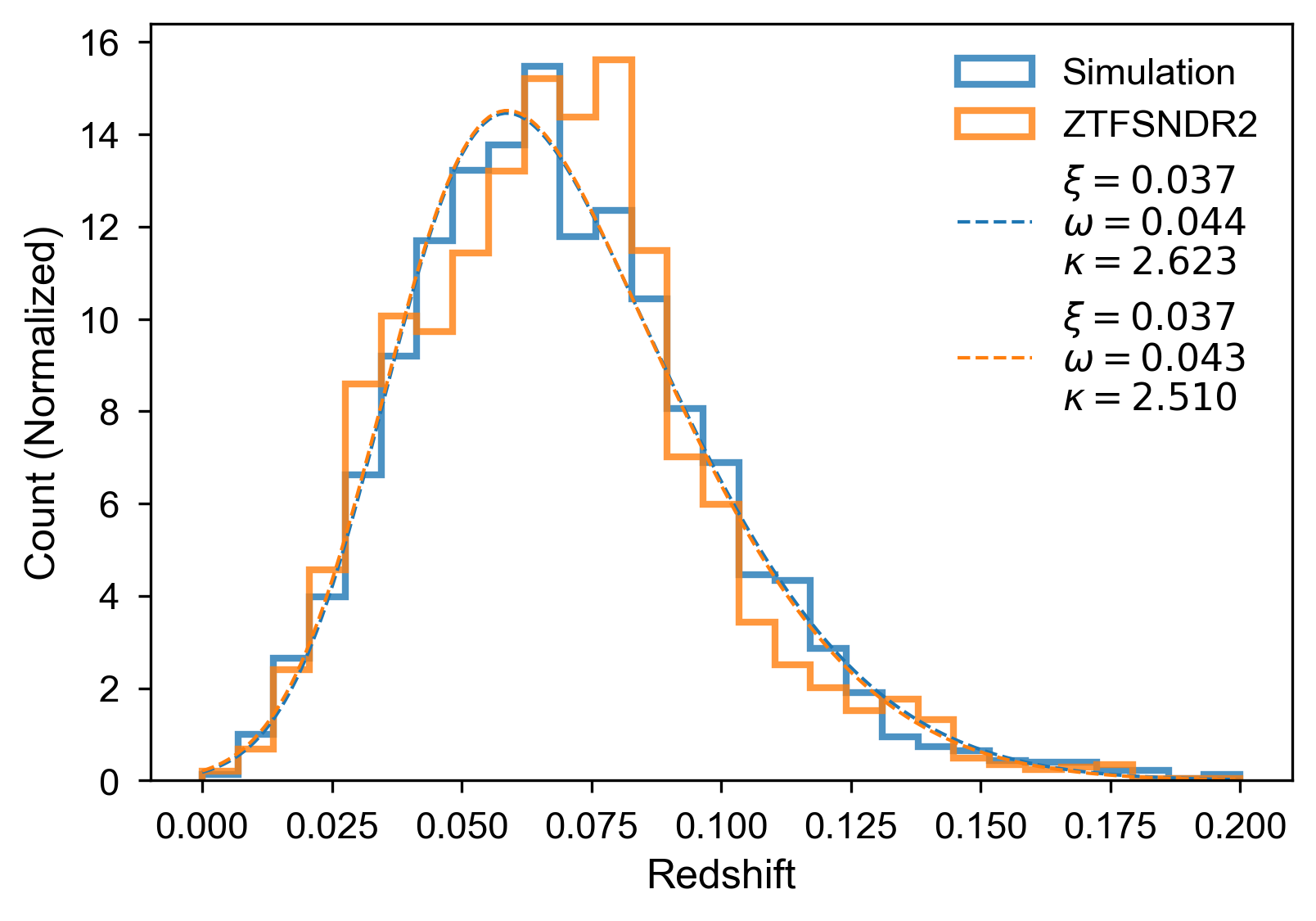}
        \caption{The redshift distributions of the simulation (blue histogram) and \ztfsn\ (orange histogram). 
        Both samples are after light curve quality cuts described in Section \ref{subsec:selection_effects}. 
        The dashed lines show the best-fit skewed-normal distribution (Eq.~\ref{eq:skewed_normal}). 
        The \ztfsn\ shows some deviations from the skewed-normal distribution, which is likely due to statistical fluctuations and/or unknown systematics. 
        Overall, the two distributions agree well with each other (with KLD=0.02).
        }
        \label{fig:z_distr}
    \end{figure}

    We first show the redshift distribution in Figure \ref{fig:z_distr}. 
    Both the simulation and \ztfsn\ in the figure include only SNe that pass the light curve quality cuts. 
    We fit both distributions to a skewed normal distribution,
    \begin{equation}
    \label{eq:skewed_normal}
    p(x) =
    \frac{2}{\omega \sqrt{2\pi}}
    e^{-\frac{(x-\xi)^2}{2\omega^2}}
    \int_{-\infty}^{\kappa
    \left(\frac{x-\xi}{\omega}\right)}
    \frac{1}{\sqrt{2\pi}}
    e^{-\frac{t^2}{2}}
    \, dt
    \end{equation}

    and report the best-fit parameters in the figure.
    Overall, the best-fit parameters agree well with each other (with $\Delta \xi = -0.0001 \pm 0.0015$,
$\Delta \omega = 0.0007 \pm 0.0016$,
and $\Delta \kappa = 0.11 \pm 0.25$). 
    We also computed the Kullback-Leibler Divergence (KLD) from the simulated distributions to the \ztfsn\ distributions $D_{\mathrm{KL}}(p_{\rm data}\|p_{\rm sim})$.\footnote{
    The Kullback-Leibler divergence $D_{\mathrm{KL}}(P || Q) = \sum_{i} P(i) \log \frac{P(i)}{Q(i)}$ quantifies the information loss when the distribution Q is used to approximate the reference distribution P.  Note that the KLD is asymmetric; $D_{\mathrm{KL}}(P \| Q) \neq D_{\mathrm{KL}}(Q\| P)$.
    }
    The KLD value is 0.02, indicating good agreement.
    The \ztfsn\ data shows some fluctuations around z=0.05, likely due to statistical fluctuations or unknown systematic effects that are not modeled in the simulation.
    
    \paragraph{SALT parameter distributions}
    
\begin{figure}
    \centering
    
    \begin{subfigure}[b]{0.48\linewidth}
        \centering
        \includegraphics[width=\linewidth]{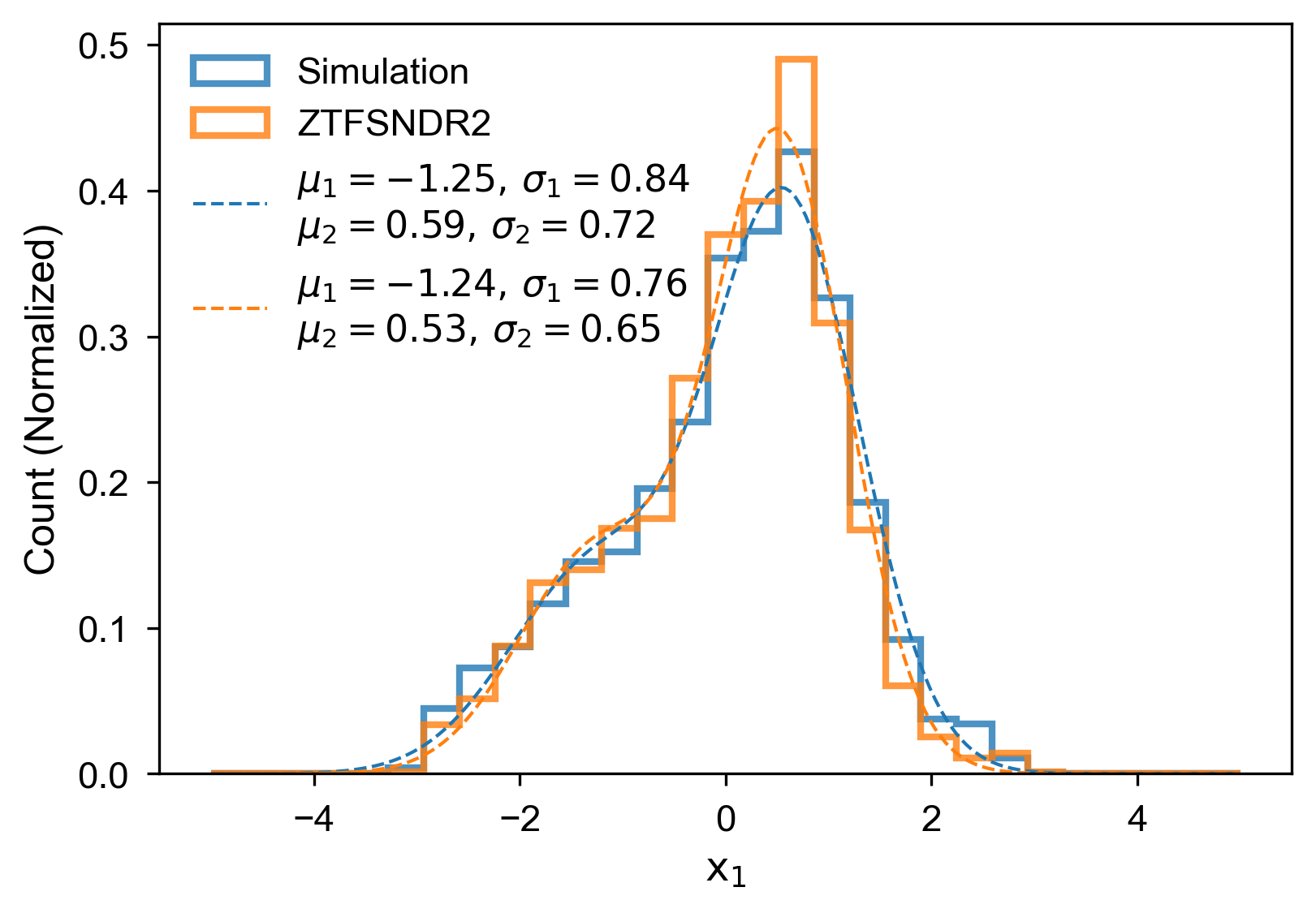}
        \caption{The $x_1$ distributions.}
        \label{fig:x1_distr}
    \end{subfigure}
    \hfill
    \begin{subfigure}[b]{0.48\linewidth}
        \centering
        \includegraphics[width=\linewidth]{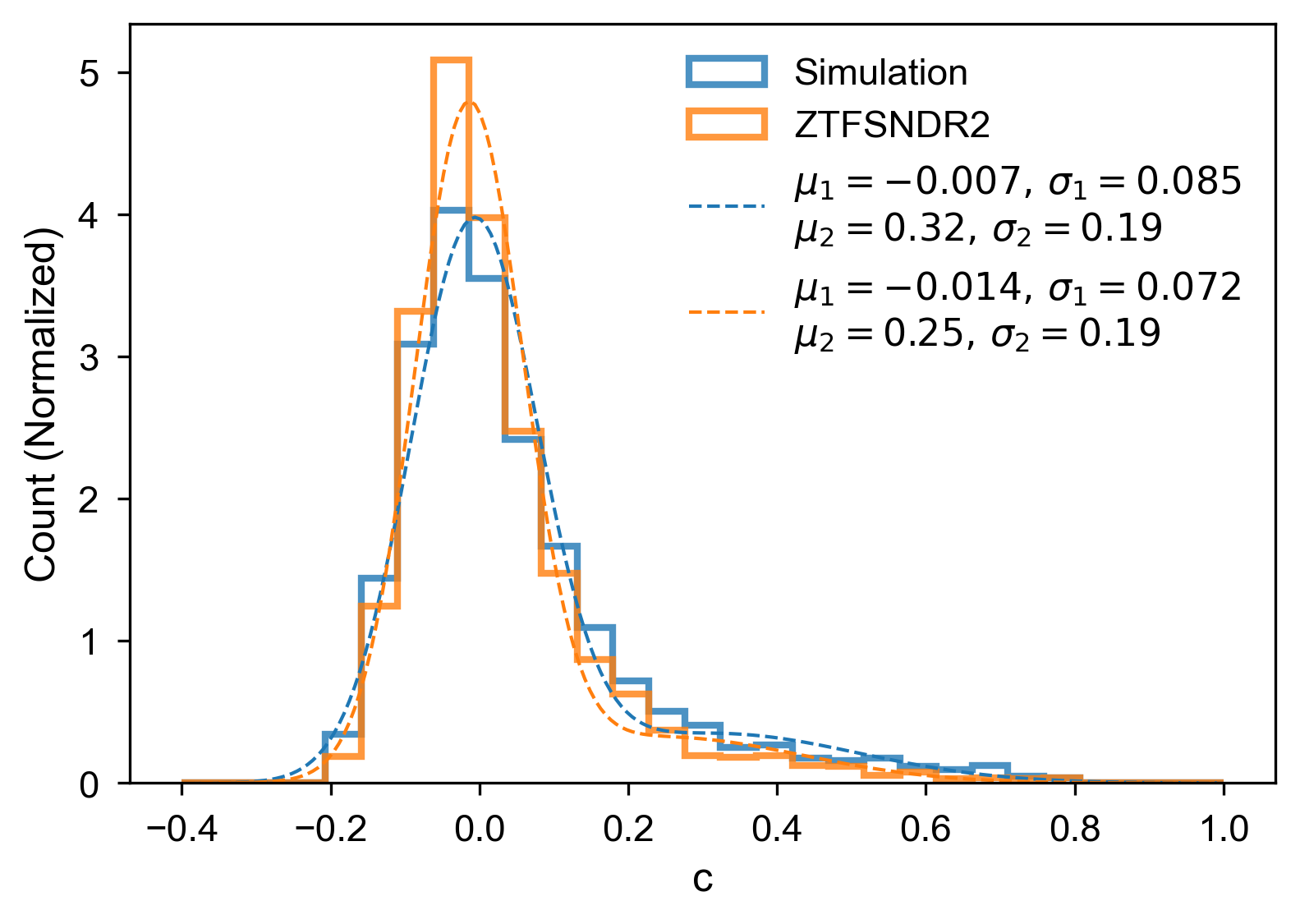}
        \caption{The $c$ distributions.}
        \label{fig:c_distr}
    \end{subfigure}

    \caption{Comparison of the SALT parameters $x_1$ and $c$ distributions for the simulation and \ztfsn\ after light curve quality and light curve fitting cuts. 
    For both plots the blue histograms are the $x_1$(a)/$c$(b) parameter distributions of the simulated values, and the orange histograms are the $x_1$(a)/$c$(b) parameter distributions of the \ztfsn\ values. 
    The dashed lines show the best-fit Gaussian Mixture models with 2 components for the simulations (blue) and \ztfsn\ (orange) respectively. 
    The simulated distributions agree well with the \ztfsn\ distributions (with KLD[$x_1$]=0.013, and KLD[$c$]=0.025.)}
    \label{fig:x1_c_distr}
    \end{figure}

    We compare the fitted $x_1$ and $c$ distributions with \ztfsn\ in Figure \ref{fig:x1_distr} and \ref{fig:c_distr}. 
    Since the light curve fitting results are meaningful only if the light curves pass both the light curve quality cuts and the light curve fitting cuts, we include here the simulation that pass the light curve quality cuts and a series of light curve fitting cuts (Table \ref{tab:sn_cuts}), and filter the ZTF results for SNe that pass the same cuts. 
    We have also confirmed that the fitted light curve parameters are consistent with the simulated values. 
    Our simulation produces similar distribution as we see in the data. 
    We fit both the simulated and data distributions with a Gaussian Mixture model and report the best-fit values in the figures. 
    The KLD values for $x_1$ and $c$ are 0.01 and 0.02, respectively.

    \paragraph{Host mass distribution and $\textrm{host mass}-x_1$ relation}

\begin{figure}
        \centering
        \includegraphics[width=0.5\linewidth]{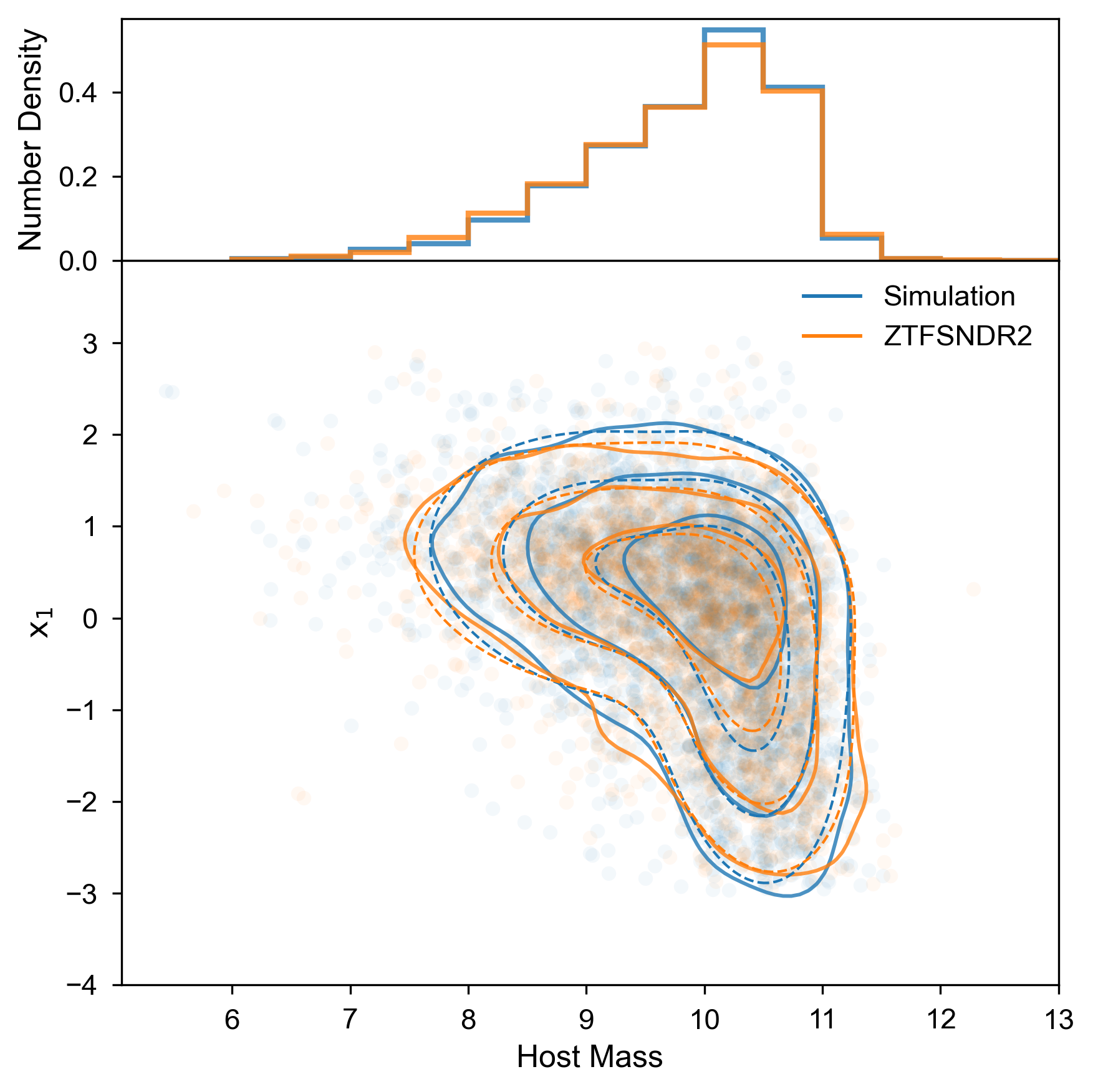}
        \caption{SALT $x_1$ parameter vs host galaxy stellar mass. 
        Main (Bottom) panel: 
        Solid lines show the density contours of the Host Mass - $x_1$ space, estimated using Kernel Density Estimation (KDE); 
        dashed lines show the best fit Gaussian Mixture model contours. 
        Blue lines represent the simulation; 
        orange lines represent \ztfsn. 
        Scatter plot of individual Host Mass - $x_1$ pairs are shown in transparent dots. 
        Top panel: 
        the host galaxy mass distribution for the simulation (blue) and \ztfsn\ (orange). 
        Following the data-driven approach in Section \ref{subsubsec:pzflow}, the simulation closely resembles the \ztfsn\ data.}
        \label{fig:host_x1}
    \end{figure}

    Here we test our data-driven approach for generating relations between SN host mass and the $x_1$ parameter. 
    The relation between host mass and $x_1$ has been reported in previous studies \citep{Sullivan2010,Johansson2013,Ginolin2025}. 
    We show that our approach in Section~\ref{subsubsec:pzflow} produces similar host mass distributions and similar host mass - $x_1$ relation (Figure \ref{fig:host_x1}) compared to the data. 
    The light curve fitting cuts are applied to both the simulation and the data to obtain a well-constrained surface.

\subsubsection{Noise Properties}

Our noise model is described in Section \ref{subsec:noise}. 
We focus on comparing two properties: 
the signal-to-noise ratio and the flux and flux error contours.

    \paragraph{Signal-to-noise ratio}
    
\begin{figure}
    \centering
    \begin{subfigure}[t]{0.48\linewidth}
        \centering
        \includegraphics[width=\linewidth]{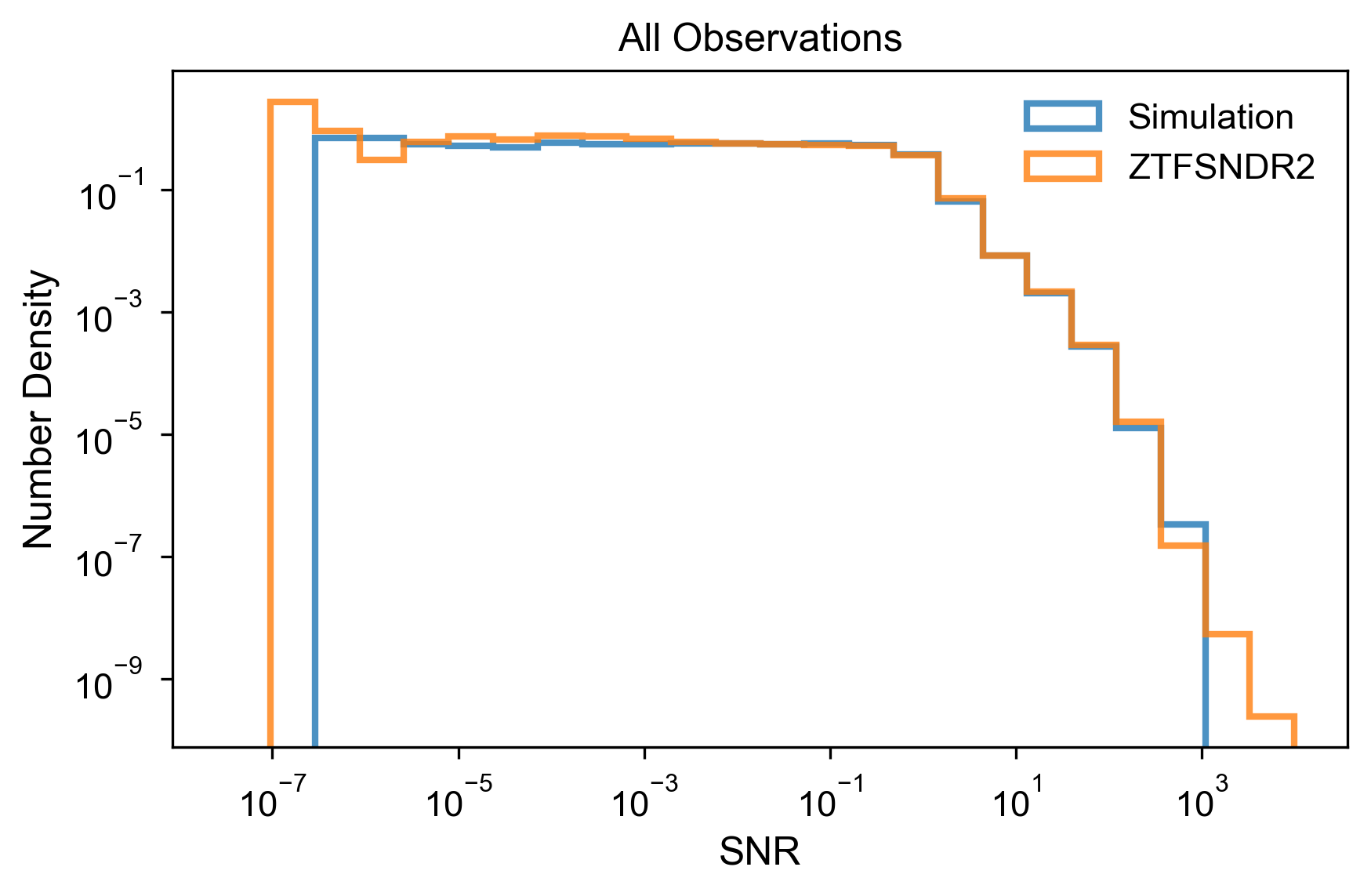}
        \caption{SNR distributions for all the observations in the simulation (blue) and \ztfsn\ (orange). 
        The y-axis shows normalized number density in log scale. 
        The simulated distribution matched \ztfsn\ for SNR up to $\sim$ 1000. 
        The simulation does not produce SNR $> 1000$ as seen in the \ztfsn\ data. 
        This is likely due to contributions from very bright objects (beyond nominal SN brightness), or under extreme observing conditions, that are not included as part of the simulation.}
        \label{fig:snr_allobs}
    \end{subfigure}
    \hfill
    \begin{subfigure}[t]{0.48\linewidth}
        \centering
        \includegraphics[width=\linewidth]{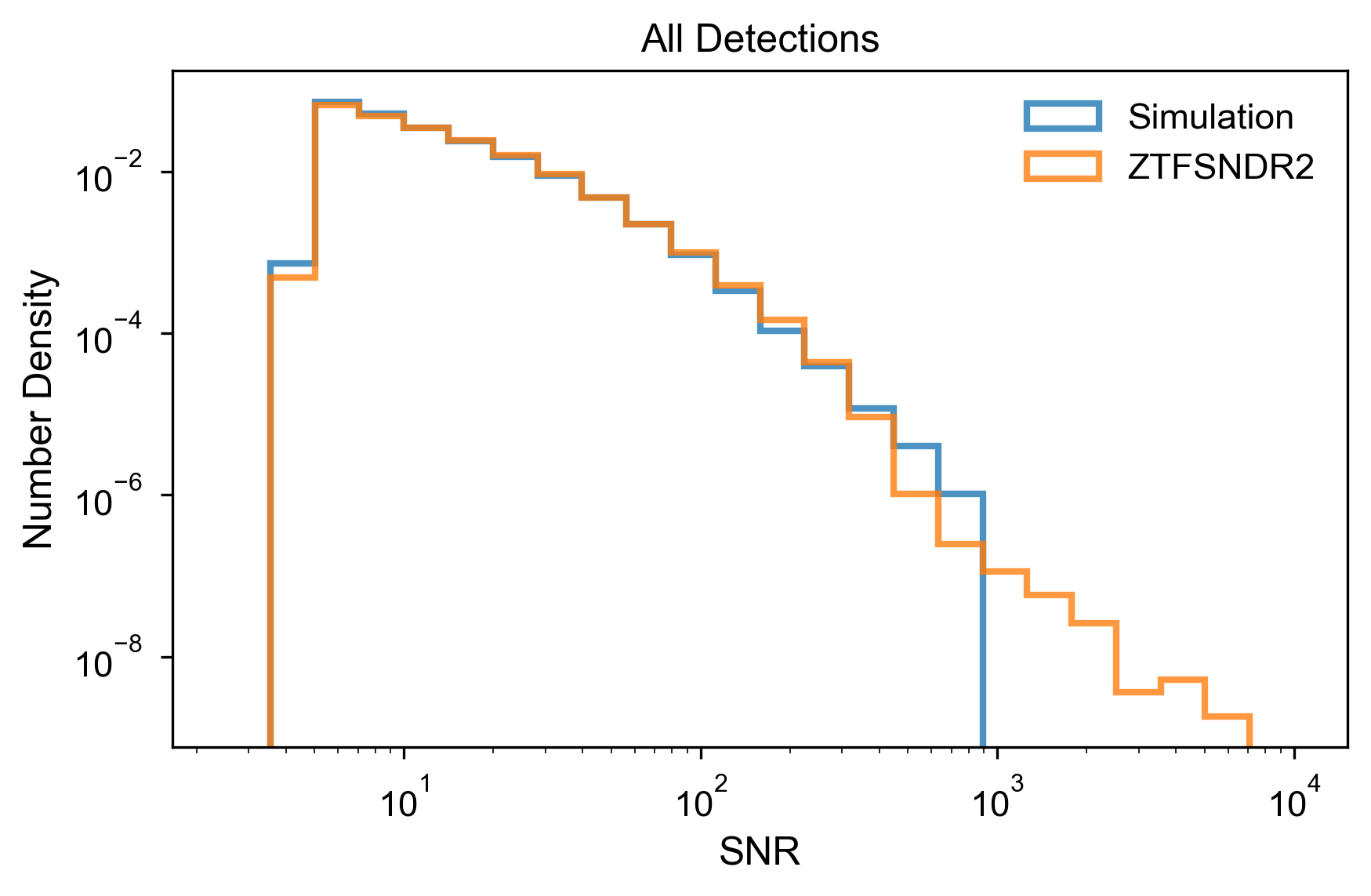}
        \caption{SNR distributions for all the detections (SNR $> 5$) in the simulation (blue) and \ztfsn\ (orange). 
        The y-axis shows normalized number density in log scale. 
        The simulated distribution matched \ztfsn\ for SNR up to $\sim$ 400. 
        Same as in Fig \ref{fig:snr_allobs}, the simulation does not produce objects with SNR $> 1000$ seen in the \ztfsn\ data.}
        \label{fig:snr_alldetection}
    \end{subfigure}

    \caption{Signal-to-noise-ratio (SNR) comparisons between the simulation and \ztfsn.}
    \label{fig:snr_comparison}
\end{figure}

    We compare the signal-to-noise-ratio (SNR) distributions for all observations for the spectroscopic sample, and for detections only, shown in Figure \ref{fig:snr_allobs} and \ref{fig:snr_alldetection}. 
    For both observations and detections, the SNR distributions are closely aligned except for the high SNR tails. 
    The maximum SNR in the simulation is around 1000, which roughly equals to a nominal SNR at $\sim 13$ mag. 
    The \ztfsn\ data contain a small fraction of observations with SNR $> 1000$, which may be the effects of very bright objects or extreme observing conditions in the image that are not modeled in our simulation.
    
    \paragraph{Flux vs flux error contours}
    
\begin{figure}
    \centering
    \begin{subfigure}[t]{0.48\linewidth}
        \centering
        \includegraphics[width=\linewidth]{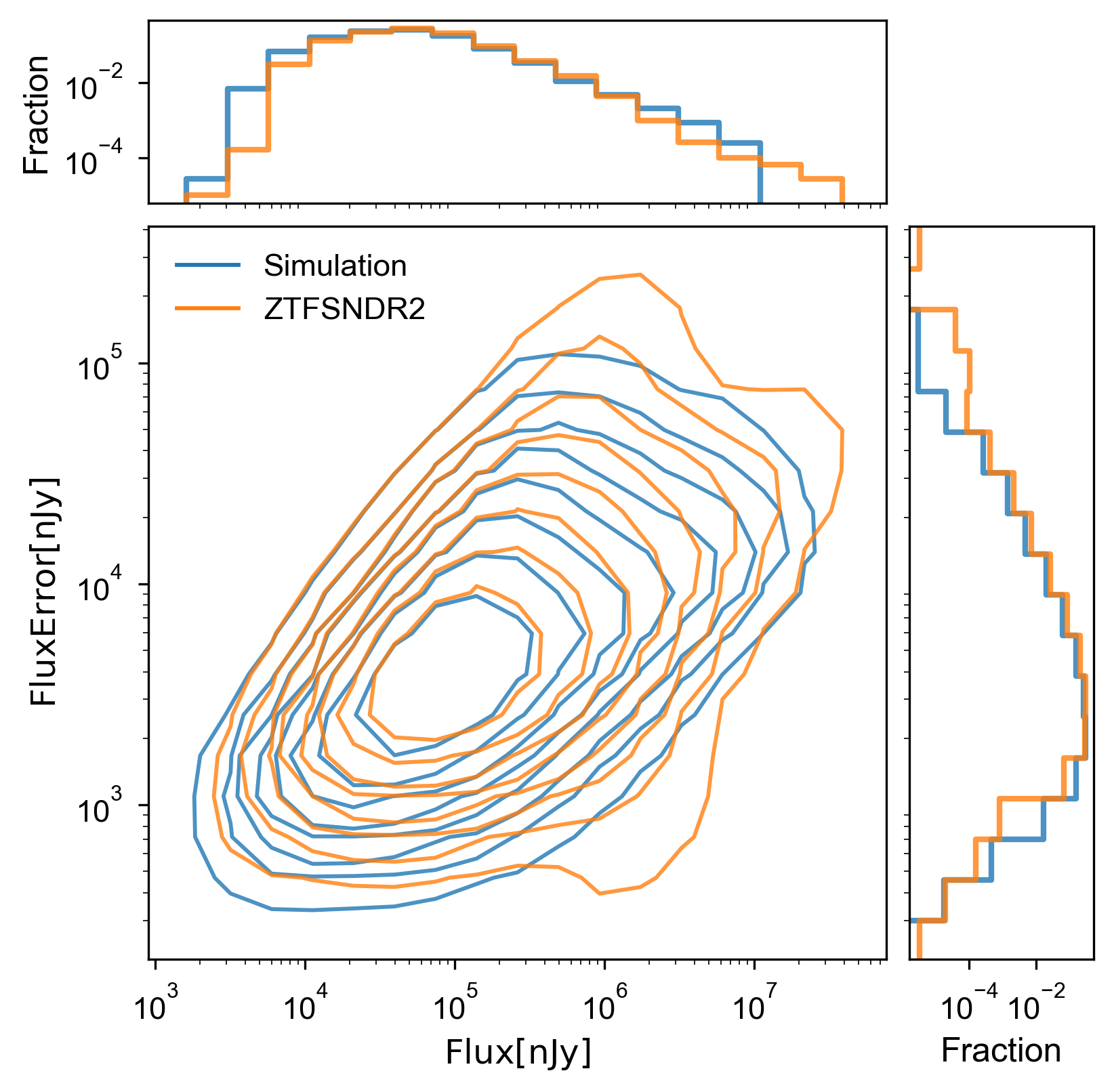}
        \caption{Flux and flux error contours for simulation (blue) and \ztfsn\ (orange). 
        The contours are fractions of counts (counts/total counts) in each log bin at the level of $10^{-1.5}$, $10^{-2}$, $10^{-2.5}$, $10^{-3}$, $10^{-3.5}$, $10^{-4}$, $10^{-4.5}$ and $10^{-5}$ from the innermost to the outermost; 
        the contours are smoothed using a Gaussian kernel with sigma=1. 
        The histograms on the top and right are also fractions of the flux, and the flux error, respectively. 
        The simulation produced similar contours to the \ztfsn\ data, with slightly larger fractions of low flux and low flux errors.}
        \label{fig:logflux}
    \end{subfigure}
    \hfill
    \begin{subfigure}[t]{0.48\linewidth}
        \centering
        \includegraphics[width=\linewidth]{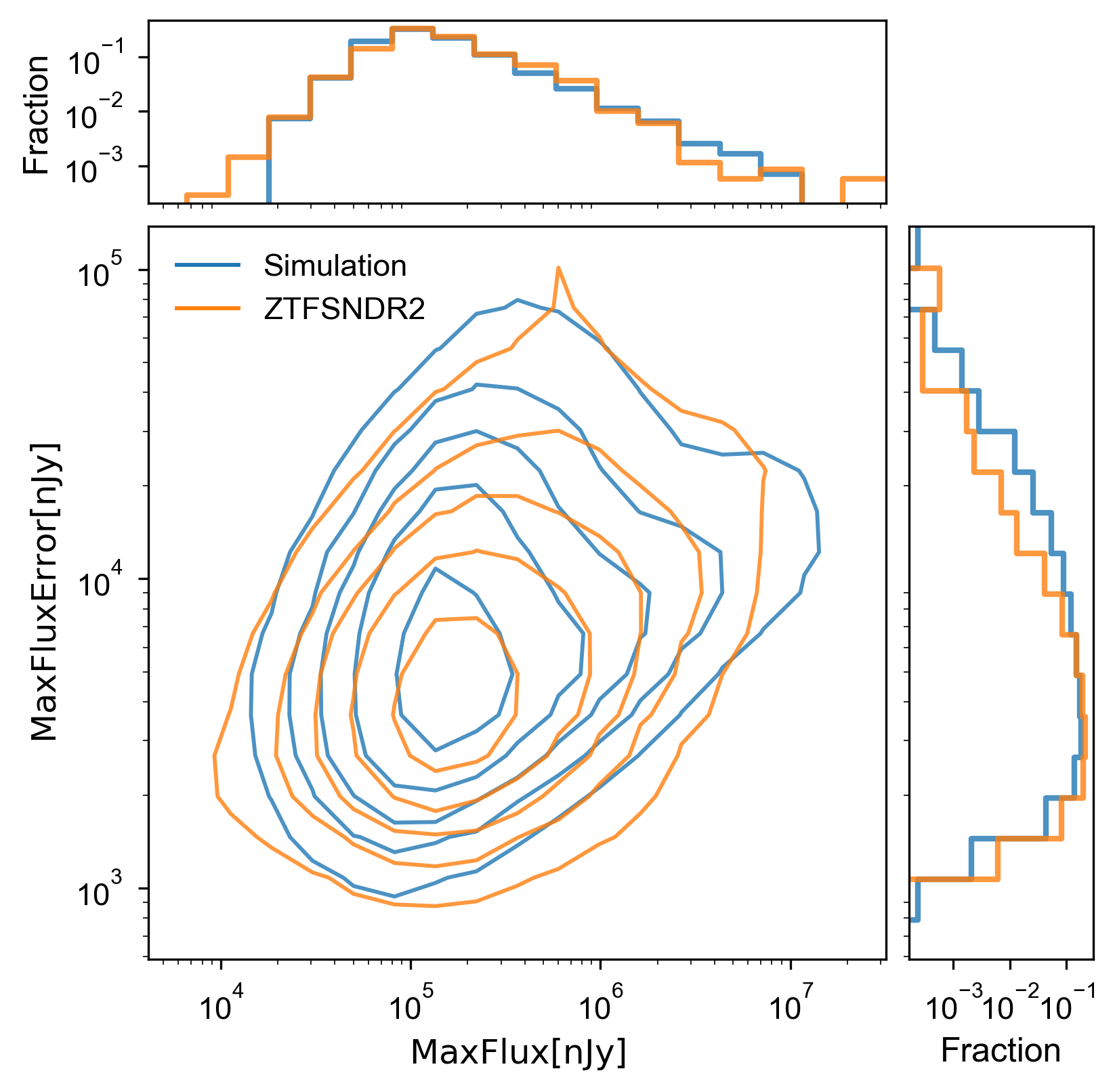}
        \caption{Flux and flux error contours at the maximum flux for simulation (blue) and \ztfsn\ (orange). 
        The contours are fractions of counts (counts/total counts) in each log bin at the level of $10^{-1.5}$, $10^{-2}$, $10^{-2.5}$, $10^{-3}$ and $10^{-3.5}$ from the innermost to the outermost; 
        the contours are smoothed using a Gaussian kernel with sigma=1. 
        The histograms on the top and right are also fractions of the flux, and the flux error at the maximum flux, respectively. 
        The simulation produced similar contours to the \ztfsn\ data.}
        \label{fig:logmaxflux}
    \end{subfigure}

    \caption{Flux and flux error contours for all detections (a), and all detections at the maximum flux of each SN (b).}
    \label{fig:logflux_combined}
\end{figure}

    We compare the flux vs flux error contours for all detections in Figure \ref{fig:logflux}, and the maximum flux - flux error contours at the maximum fluxes in Figure \ref{fig:logmaxflux}. 
    In Fig \ref{fig:logflux}, both the fluxes and flux errors have slightly higher fractions of low values, which leads to a larger tail in the bottom-left corner of the plot. 
    In Fig \ref{fig:logmaxflux}, we observe that the flux distribution at the epoch of the maximum flux are shifted slightly toward the lower flux value side compared to the data, while the flux error distribution at the same epoch are shifted slightly to the higher flux error side. 
    The small differences may be due to the light curve model not fully representing the data, and/or as we have discussed in Section \ref{subsec:obslog}, the uncertainty in the error estimation. 
    Overall, these distributions and contours are very similar. 
    
\subsection{Hubble Diagram}\label{subsec:hubble_diagram}

\begin{figure}
    \centering
    \includegraphics[width=0.5\linewidth]{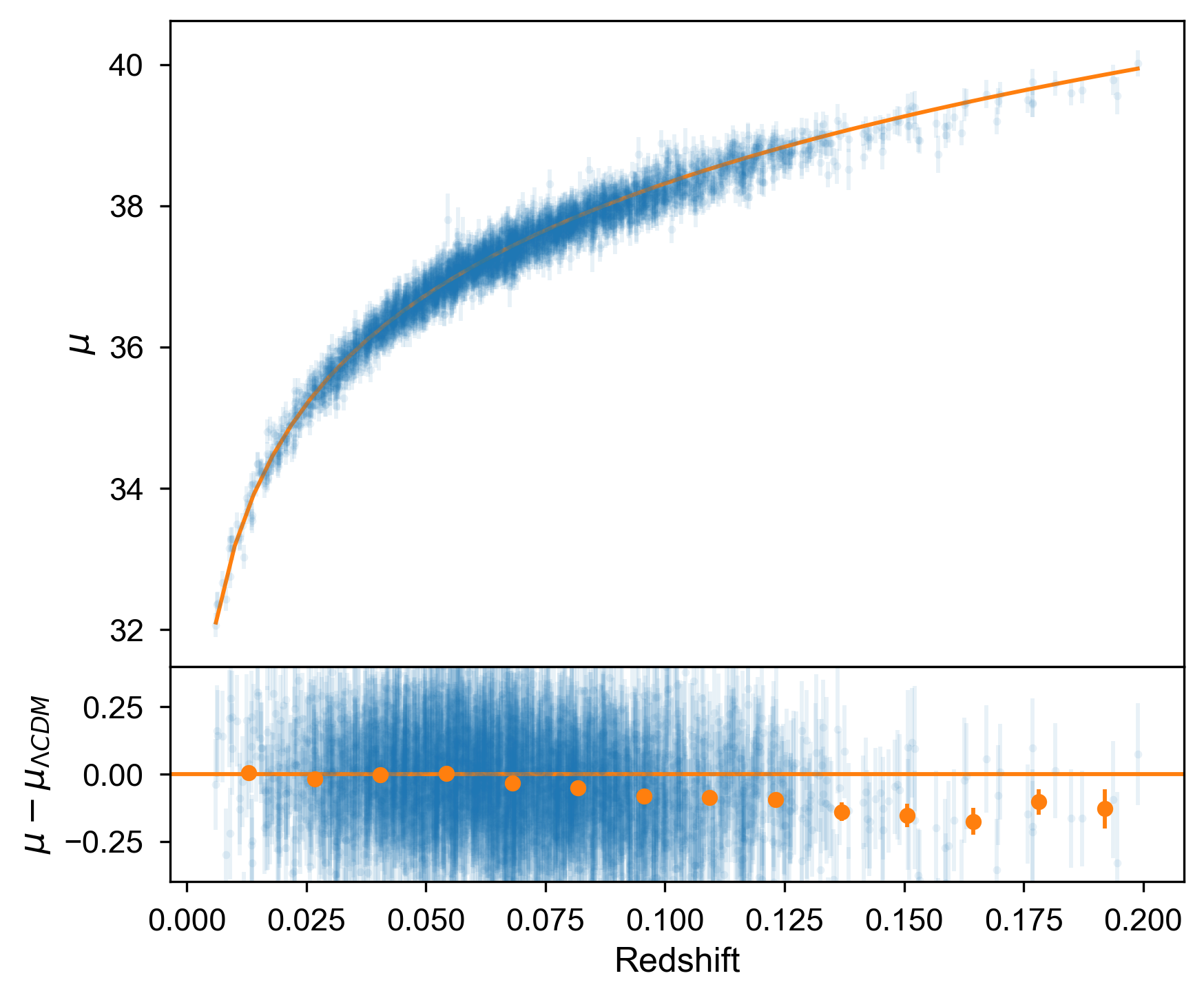}
    \caption{The Hubble Diagram from the simulated sample. 
    Upper panel: 
    Hubble diagram of 2992 simulated SN Ia that passed the selection and quality cuts defined in Table \ref{tab:sn_cuts}. 
    The blue points are the distance modulus of each individual SN (Eq.~\ref{eq:tripp}); 
    the orange curve is the distance modulus calculated using $\Lambda CDM$ cosmology. 
    Lower panel: 
    The Hubble residuals (HR $=\mu - \mu_{\rm \Lambda CDM}$) as a function of redshift. 
    Blue points are the HR of each individual SN; 
    orange points represent means and errors on the means of the binned Hubble residuals in 15 redshift bins. 
    No bias corrections are accounted for when computing the distance modulus. 
    As expected, we observe a selection bias in HR starting at $z \sim 0.06.$}
    \label{fig:HD}
\end{figure}

We compute the distance moduli using the Tripp formula (Eq.~\ref{eq:tripp}) and the cosmological parameters used for simulations. 
The Hubble diagram is shown in Figure~\ref{fig:HD}. 
Without applying any bias correction term, we find that the Hubble residuals ($\mu-\mu_{{\Lambda}\textrm{CDM}}$) are biased toward negative values starting around $z=0.06$.
This is consistent with the results in \cite{Amenouche2025}, which claims the sample is free from selection bias up to $z=0.06$. 
The RMS of the Hubble residuals is 0.18, which is consistent with both the data and the input scatter that we used for simulation.

\begin{figure}
    \centering
    \includegraphics[width=0.5\linewidth]{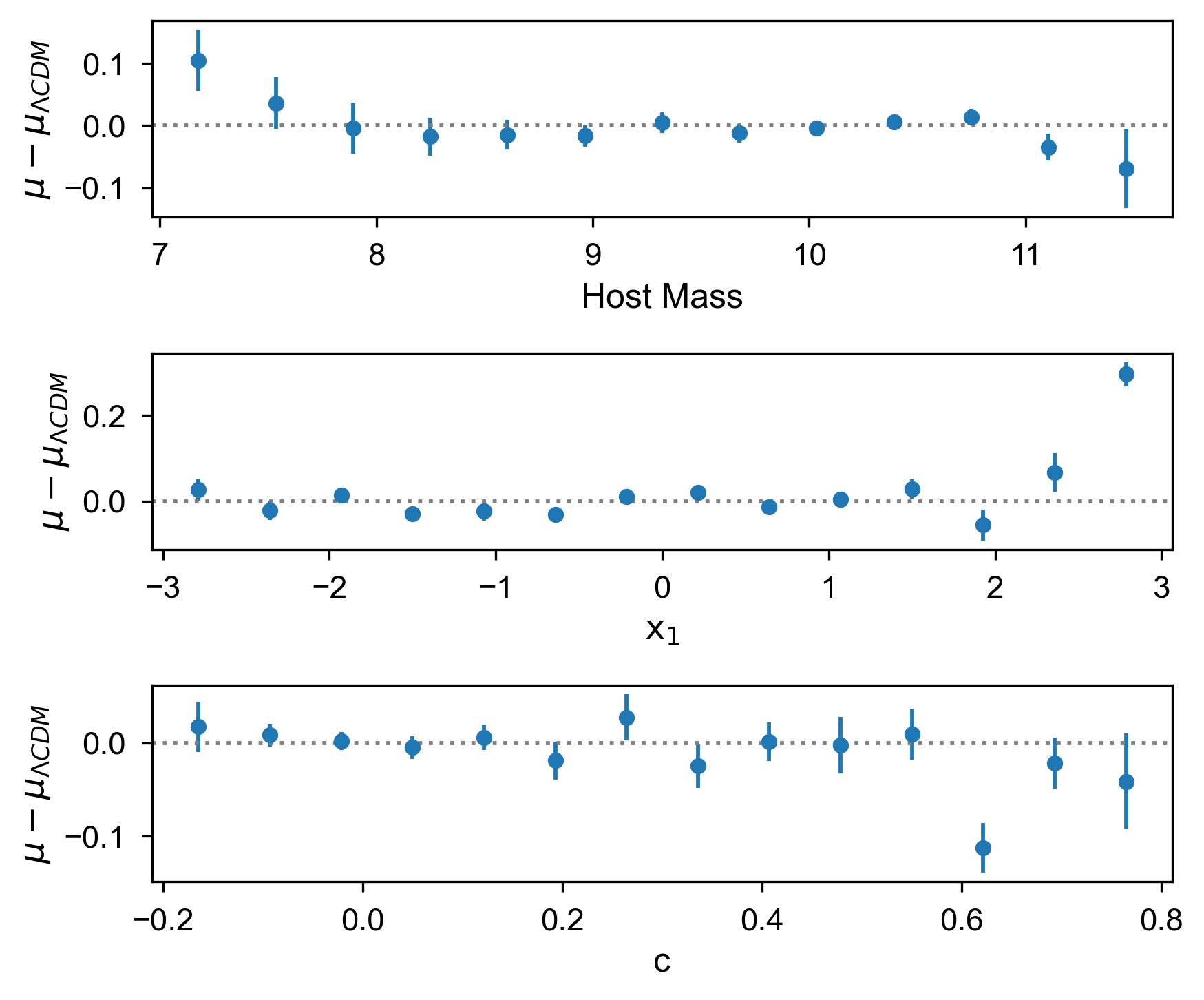}
    \caption{Binned Hubble residuals against host galaxy mass, and light curve parameters $x_1$ and $c$ for a sample with $z<0.06$. 
    No significant bias is present in any of these parameter spaces.}
    \label{fig:mures_par}
\end{figure}

Although we did not simulate any relations between the Hubble residuals and the host galaxy mass (for example, a mass step), we would like to check if any systematics arise which may be due to imperfect linear corrections for the $x_1$ or $c$ parameters. 
We show in Fig.~\ref{fig:mures_par} the binned Hubble residuals for the volume-limited ($z<0.06$) sample. 
The binned Hubble residuals are consistent with 0 in most of the bins, indicating no significant systematics. 
There are slight deviations in regions where $x_1 > 2$ and $c>0.5$, which may require future investigation.

\subsection{Simulation Performance}

The simulation is generated on a Macbook Pro with an Apple M1 Max chip (10-core CPU, 32-core GPU) with 64~GB of memory. 
We generated a total of 84,688 SNe with 75,365,656 observations. 
The average SN generation rate is $\sim 100$ SN per second, about 10 ms per SN. 
Utilizing \lynx's built-in parallelization features and using 8 jobs and a batch size of 5000, the total CPU time is $\sim 26.4$~s, plus $\sim 4.5$ minutes startup overhead. 
The observing log contains 522,192 rows and 24 columns; 
the load-in time is $\sim 500$ ms.

\section{Discussions and Conclusion}\label{sec:discussion}

In this paper, we introduce \lynx, a Python-based software framework for forward modeling transient and variable light curves with realistic observing conditions and uncertainties. 
\lynx\ provides a flexible, user-friendly, and extensible framework that meets the growing needs of the time-domain astronomy community in the era of large and deep surveys such as LSST and Roman.

We demonstrate the functionalities and validate the credibility of \lynx\ through a use case of simulating a realistic SN Ia sample as observed by ZTF. 
We show that users have the flexibility to define a set of input parameters and their distributions, and generate simulations directly from an observing log (typically provided by the survey). 
For SN Ia, we generate the simulation using a realistic SN Ia SED model, volumetric rate and luminosity function. 
We model the SN Ia light curve parameter distribution, the host galaxy stellar mass distribution and selection functions using a data-driven approach. 
We use realistic detector characterizations and the publicly-available observing log to generate realistic cadence and noise. 

We compare our simulation to the \ztfsn\ data release. 
The resulting redshift distribution, SN Ia light curve parameter distributions, host galaxy mass distribution all match the data very well. 
We also demonstrated that we can model joint parameter distributions using a data-driven approach utilizing \pzflow. 
The sample statistics in our simulation match the data at various stages of the analyses (e.g, applying a spectroscopic efficiency, applying data quality cuts and light curve fitting quality cuts.) 
The use of \sncosmo\ and \pzflow\ in our simulation demonstrates how the modular structure of \lynx\ allows users to easily incorporate community packages into the simulation flow to make use of previously implemented models and modeling frameworks. 
Finally we show that we reproduce similar statistics in the Hubble diagram and confirmed the sample completeness up to $z=0.06$, consistent with \cite{Amenouche2025}.

Since we simulate an existing survey, an accurate and detailed observing log that contains necessary information such as zero points, seeing, sky background/sky noise, and accurate instrument characterizations are the key to reproducing realistic cadence and uncertainties. 
We encourage survey operators to make such information available with detailed documentation to support future analyses and forecasting. 

This paper also serves a working example of how to build a simulation pipeline from scratch. 
Users can build their own simulation pipeline by adding and/or replacing parameter models, physical models, observational effects, survey data, etc. 
With detailed and realistic simulations of SN Ia like this, we can test assumptions such as dependency on host galaxy properties, effect of parameter evolutions over redshift, or biases in various analysis stages, for example. 
Our efficient simulation strategy utilizing parallel computing makes it natural to explore more computational extensive approaches such as simulation-based inference, which is left for future studies.

We confirm that \lynx\ is ready to be used by the community, and we encourage community feedback and contributions moving forward.

\appendix
We describe the caveats in calculating sky background in Appendix \ref{appx:skybg}.\\

\section{Caveats about Sky Background}\label{appx:skybg}

The sky background is needed to simulate realistic noises. While we intended to utilize public information regarding sky background from the metadata DB,
the values provided in the metadata DB do not seem to represent the actual sky background in the images;
they are instead a ``robust estimate of background level in scaled science image," (possibly as a result of rescaling the science image to the reference image).
In order to obtain a representative sky background for realistic noise estimation, we derived the sky background utilizing existing information using Eq.~\eqref{eq:noise}, assuming that the 5$\sigma$ magnitude limit is the magnitude at which $\texttt{flux}/\texttt{flux error}= 5$. 
Note that there are two limiting magnitude columns in the observing log, provided by both \cite{Rigault2025} and the metadata DB. 
We noticed these two columns are not identical; 
the \texttt{maglimit} column in the observing log is described as the 5$\sigma$ limiting magnitudes of the science images, while the \texttt{maglim} column in the metadata DB is described as the ``Median of IMQA Records of Magnitude limit of PSF-fit catalog based on semi-empirical formula [mag].'' 
We experiment using both columns to derive different sky background values and generated simulations, respectively. 
Interestingly, when comparing the simulation and the corresponding \ztfsn\ data, we find that using the observing log value leads to underestimating the flux errors by 12\%, while using the metadata DB values leads to overestimating the flux errors by 19\%. 

In a similar effort, \cite{Amenouche2025} compared the simulated errors in two simulations:
one used a sky noise value derived from the limiting mag from the science images and the other used that derived from the difference images.
They concluded that the science image values generate underestimated errors that are nonetheless closer to those reported in the data and that the difference image mag limit is not correctly estimated. 
While we do not have easy access to the difference image limits, as they are not included in the \ztfsn\ data release, the underestimation we observe when using the science image limits is consistent with what is found in \cite{Amenouche2025}. 
We choose to use the 5$\sigma$ magnitude limit published by the \ztfsn\ data for the results of this paper.

\begin{acknowledgments}
LINCC Frameworks is supported by Schmidt Sciences. A.C. acknowledges support from the DiRAC Institute in the Department of Astronomy at the University of Washington. The DiRAC Institute is supported through generous gifts from the Charles and Lisa Simonyi Fund for Arts and Sciences, and the Washington Research Foundation. This work made use of \textit{Claude (Anthropic)} and \textit{ChatGPT (OpenAI)} for partial editing of the manuscript text.
\end{acknowledgments}

\facilities{PO:1.2m, PO:1.5m}

\software{
\texttt{Astropy} \citep{astropy:2013, astropy:2018, astropy:2022},
\texttt{dust\_extinction} \citep{Gordon2024}
\lynx\ (\citealt{lynx_software}; Dai et al. 2026, this work),
\texttt{Matplotlib} \citep{Hunter2007}, 
\texttt{Numpy} \citep{Harris2020},
\texttt{Nested-Pandas} \citep{nested_pandas},
\texttt{Pandas} \citep{reback2020pandas,mckinney-proc-scipy-2010}, 
\pzflow\ \citep{Crenshaw2024},
\sncosmo\ \citep{Barbary2016,barbary_2025_15019859},
\texttt{scikit-learn} \citep{scikit-learn},
\texttt{Scipy} \citep{Virtanen2020},
\texttt{sfdmap2} \citep{sfdmap2},
}

\bibliography{ztfsn}{}
\bibliographystyle{aasjournalv7}

\end{document}